\documentclass[conference]{IEEEtran}
\IEEEoverridecommandlockouts
\usepackage{cite}
\usepackage{subcaption}
\usepackage{amsmath,amssymb,amsfonts}
\usepackage{algorithmic}
\usepackage{graphicx}
\usepackage{textcomp}
\usepackage{xcolor}
\usepackage{booktabs}
\usepackage{float}

\def\BibTeX{{\rm B\kern-.05em{\sc i\kern-.025em b}\kern-.08em
    T\kern-.1667em\lower.7ex\hbox{E}\kern-.125emX}}
\begin{document}

\title{Who Speaks What from Afar: Eavesdropping In-Person Conversations via mmWave Sensing
}

\author{
\IEEEauthorblockN{Shaoying Wang\IEEEauthorrefmark{1},
Hansong Zhou\IEEEauthorrefmark{1},
Yukun Yuan\IEEEauthorrefmark{2},
Xiaonan Zhang\IEEEauthorrefmark{1}}
\IEEEauthorblockA{\IEEEauthorrefmark{1}Florida State University, Tallahassee, FL, USA}
\IEEEauthorblockA{\IEEEauthorrefmark{2}University of Tennessee at Chattanooga, Chattanooga, TN, USA}
\thanks{Corresponding author: Xiaonan Zhang (\texttt{xzhang14@fsu.edu}).}
\vspace{-0.2in}
}

% \and
% \and
% \and
% \and
% \and

\maketitle

\begin{abstract}
Multi-participant meetings, such as business or medical discussions, often involve sensitive content. Understanding both the speech content and the speaker attribution is essential for meaningful interpretation. Prior methods leverage supervised learning to reconstruct high-frequency speech features, which can help discriminate speakers. However, they rely on paired high-quality audio data for training, which is not always available.
In this paper, we design an unsupervised mmWave-based eavesdropping attack that answers the question “who speaks what?” By leveraging the spatial diversity introduced by ubiquitous objects, the attacker can remotely capture in-person conversations without requiring clean speech audio or speaker identities. Specifically, we extract speaker-specific patterns for speaker distinction by amplifying weak frequency responses.
Meanwhile, we enhance signal quality by fusing vibration signals from multiple objects through a deep learning framework.
Our attack achieves a  success rate of 0.99 in speaker distinction and delivers consistent signal enhancement performance across diverse real-world settings, including varying sensing distances, object materials, and speaker arrangements.

\end{abstract}

\section{Introduction}
Sound-based eavesdropping poses a significant privacy threat, due to the often unencrypted nature of voice. During meetings, participants frequently discuss sensitive topics such as bid prices, confidential client information, and strategic company decisions. If intercepted, such conversations could result in serious competitive disadvantages, financial losses, or violations of confidentiality agreements. To mitigate the risk of sound leakage, many people  opt to have meetings in soundproof rooms. Nevertheless, sound waves emitted by speakers propagate through the air and induce micrometer-level vibrations on the surfaces of surrounding objects~\cite{jiang2020mmvib}. By exploiting the penetration capability of mmWave signals, an attacker outside the room can capture these subtle vibrations and potentially reconstruct the original speech content.

% Sound eavesdropping is a major privacy concern due to the unencrypted nature of voice communication. During a meeting, participants often take turns discussing sensitive topics like bid prices, confidential client information, and strategic company decisions. These conversations, if intercepted, could lead to significant competitive disadvantages, financial losses, or breaches of confidentiality agreements. To mitigate the risk of sound leakage, many people opt to conduct meetings in soundproof rooms. Nevertheless, the sound waves from speakers propagate through the air, which induce micrometer-level vibrations on the surfaces of the surrounding objects~\cite{jiang2020mmvib}. Leveraging the obstacle penetration ability of mmWave signals, these vibrations can be detected by the attacker outside the room to further infer the speech content. 

Accurate attribution of speakers is equally important as capturing the speech content in successful eavesdropping.  Misattributing a speaker can result in severe consequences. Taking the business meeting as an example, if the attacker misattributes important statement or decision to the wrong company representatives, it will result in incorrect interpretations of the company's strategy and cause financial loss. However, attributing different speakers with mmWave-based methods is hindered by its narrowband, low-SNR signal, and lossy representations of speech~\cite{wang2024vibspeech, zhang2022ambiear}, since these degraded mmWave signals lack high-frequency speech information that is essential for distinguishing speakers, such as formant details and timbre-related cues~\cite{anguera2012speaker, wang2018speaker}. Although some supervised learning~\cite{wang2022mmeve, feng2023mmeavesdropper, hu2022milliear} methods manage to recover high-frequency features from mmWave signals using paired labeled high-quality audio, this is infeasible in practical in-person conversation scenarios since the attackers generally have no access to the high-quality data.

\begin{figure}[H]
\vspace{-0.05in}
    \centering
    \includegraphics[width=0.8\linewidth]{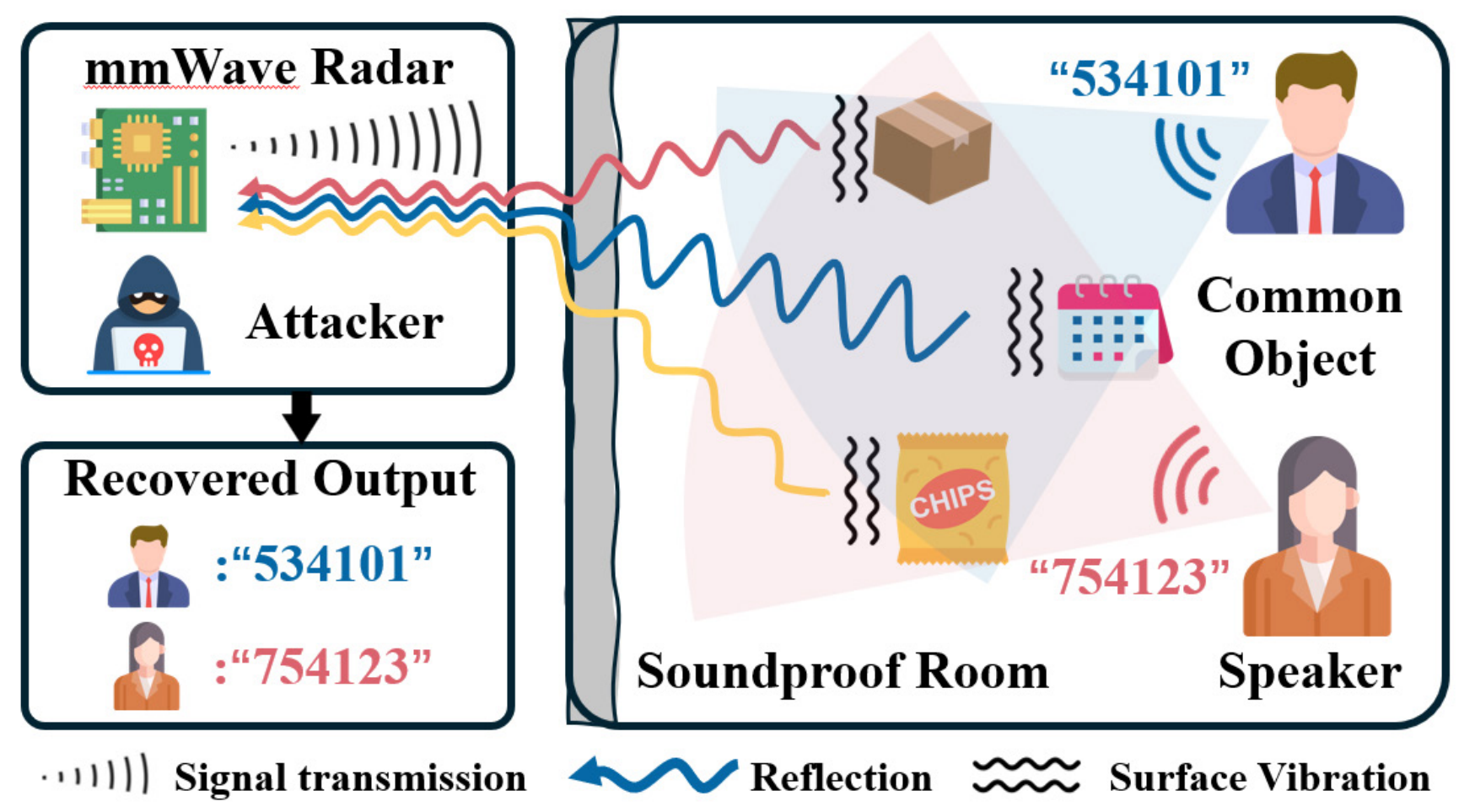}
    \vspace{-0.05in}
    \caption{Illustration of our attack scenario.}
    \vspace{-0.1in}
    % \vspace{-0.1in}
    \label{fig:multi_object}
\end{figure}

In this paper, we propose an unsupervised, multi-object mmWave-based eavesdropping attack that addresses the challenge of “who speaks what?”. Our attack requires neither clean speech audio nor speaker identities. Instead, it relies solely on mmWave signals induced by minute vibrations on multiple everyday objects in the room (e.g., paper bags, calendars, cardboard boxes). As illustrated in Fig. \ref{fig:multi_object}, the attacker collects reflected signals from multiple objects to distinguish speakers at different locations. In particular, each speaker generates unique vibration patterns on surrounding objects, reflected in their frequency responses. Even slight differences in the relative positions of speakers and objects lead to perceptible spectral differences. By analyzing these frequency responses, our attack is able to attribute individual speakers and associate them with their corresponding speech content. 

Nevertheless, implementing this attack in real world faces  critical technical challenges:
\textit{1)	Low resolution vibration:} vibration signals are extracted by isolating range bins through FFT processing. However, reflections from nearby static objects are mixed into the same bin, introducing interference that lowers the resolution of the extracted vibration signal.
\textit{2) Weak frequency response:} speech-induced vibrations often produce weak frequency responses, due to their low amplitudes. These weak responses are easily submerged by the inherent self-noise of the mmWave radar. As a result, it becomes challenging to detect speaker-specific patterns.
\textit{3)	Poor speech representation:} 
Speech attenuates during propagation due to distance and object properties. Consequently, the vibration response only retains a partial representation of the original speech, thereby degrading speech quality.

Our attack addresses these challenges as follows: \textit{1) Vibration Resolution Enhancement:}  we design a speech-aware calibration scheme to isolate speech-induced vibrations from static interference. It is based on our theoretical analysis of static interference.
% we first theoretically model the vibration interference introduced by nearby static objects. Building on this model, we design a speech-aware calibration scheme that leverages non-speech segments to effectively isolate speech-induced vibrations.
\textit{2) Frequency Response Amplification:} we use a noise-robust signal processing pipeline that amplifies speech-induced frequency responses by suppressing mmWave self-noise. 
\textit{3) Multi-object Signal Enhancement:}
we propose a deep learning framework to  improve signal quality, which aggregates complementary speech components across objects.

The contributions of the paper are summarized as follows:
\begin{itemize}
    % \item  We propose a mmWave-based in-person conversation eavesdropping attack. We are the first work that analyzes the frequency response difference among speech-induced vibrations on everyday objects for distinguishing and further enhancing the eavesdropped speech quality from different participants. 
    \item We propose a mmWave-based in-person conversation eavesdropping attack. Our attack distinguishes speech from multiple speakers by analyzing subtle speech-induced vibrations on surrounding objects.
    % \item We are the first to identify the signal drift in the presence and absence of speech, a factor that adversely impacts phase change extraction. To mitigate this, we explore a constrained circle fitting method that leverages non-speech segments to assist in calibrating the signal present in speech segments.
    \item We model the vibration interference caused by nearby static objects. A speech-aware calibration scheme is explored 
    to mitigate such interference.
    \item We develop a noise-robust signal processing pipeline for speaker distinction.  To enhance signal quality, we design a deep-learning-based framework to aggregate multi-object signals.
    % \item We deploy an effective unsupervised approach to detect speech-induced vibration differences for distinguishing participants. To enhance speech quality, we design a deep-learning-based framework that integrates signals from multiple objects by emphasizing speech components while suppressing noise components.
    \item We evaluate our attack through extensive experiments in a real-world soundproof room, demonstrating its effectiveness across various common objects, participant genders, distances, and participant positions.
\end{itemize} 

\section{PRELIMINARIES}
\subsection{Object Vibration Model}     \label{sec:vibration_model}
When an audio source emits sound, it generates pressure fluctuations in the air, which propagate as longitudinal waves, known as sound waves. These waves propagate omnidirectionally and are characterized by time-varying sound pressure \cite{jaramillo2014architectural}. The sound pressure at time 
$t$ can be represented as a superposition of sinusoidal components: 
\begin{equation}
\begin{aligned}\textstyle
P(t) = \sum\nolimits_i P_i \cos(\omega_i t + \phi_i),
\end{aligned}\label{eq:sound pressure}
\end{equation}
where $P_i$, $\omega_i$, and $\phi_i$ are the amplitude, angular frequency, and initial phase of the $i$th sinusoidal component.

\begin{figure}[t]
\centerline{\includegraphics[width=0.6\linewidth]{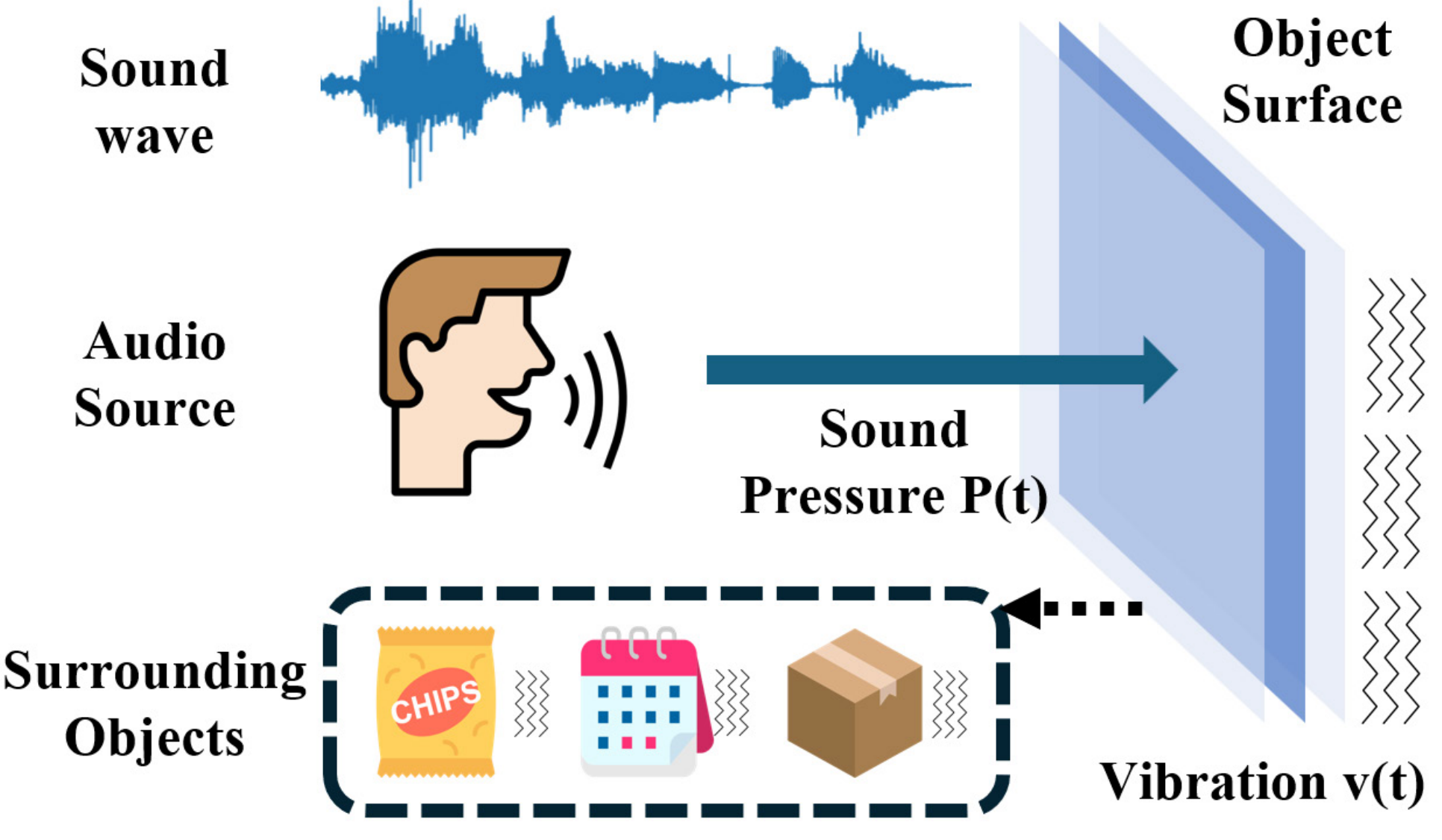}}
 \caption{Illustration of sound pressure-induced surface vibration.}
\label{fig:vibration}
\vspace{-0.2in}
\end{figure}
As shown in Fig. \ref{fig:vibration}, when sound waves reach nearby objects (e.g., chip bags, calendars, or cardboard boxes), the pressure exerts a force on their surfaces, calculated as $F = {P} \cdot {S}$,  where $S$ denotes the surface area  exposed to the wave. Due to the thin and flexible nature of these surfaces, their dynamic response can be modeled as forced vibrations of elastic membranes \cite{inbook}:
\begin{equation}
\textstyle
EI \frac{\partial^4 y}{\partial x^4} + B \frac{\partial y}{\partial t} + m \frac{\partial^2 y}{\partial t^2} = F(t),
\label{eq:pde_vibration}
\end{equation}
where \( y(x,t) \) is the transverse displacement, \( EI~(\mathrm{N \cdot m^2}) \) is the flexural modulus, \( B~(\mathrm{kg \cdot s^{-1}}) \) is the damping coefficient, and \( m~(\mathrm{kg/m}) \) is the mass per unit length.
Assuming steady-state response and spatial averaging, the resulting displacement \( d(t) \) in Eq. \ref{eq:pde_vibration} can be expressed as:
\begin{equation}
\textstyle
d(t) = \sum\nolimits_i \frac{P_i \cos(\omega_i t + \phi_i) \cdot S}{m \sqrt{(\omega_0^2 - \omega_i^2)^2 + \left( \frac{\omega_i B}{m} \right)^2}},
\label{eq:displacement_solution}
\end{equation}
where \( \omega_0 \) denotes the natural frequency of the membrane.
Eq.~\ref{eq:displacement_solution} reveals that the vibration amplitude is proportional to the local sound pressure. In free-field conditions, sound pressure decays with distance approximately as \( P(r) \propto \frac{1}{r} \) \cite{kinsler2000fundamentals}. Additionally, high-frequency components attenuate more rapidly with distance, leading to weaker induced vibrations compared to low-frequency components. Thus, closer objects experience stronger vibrations and preserve richer high-frequency content.

\subsection{mmWave-based Vibration Sensing}
We leverage the mmWave sensing based on frequency-modulated continuous-wave (FMCW) radar to capture sound-induced surface vibrations for acoustic signal reconstruction. 
Specifically, the mmWave radar emits a linearly frequency-modulated signal, commonly known as a chirp. It then receives and further modulates the reflected signal for further analysis. The transmitted and received signals, denoted as \( s_{TX}(t) \) and \( s_{RX}(t) \), are mathematically formulated as:
\begin{equation}
\begin{aligned}\textstyle
 s_{T X}(t)&=\textstyle \exp \left[j\left(2 \pi f_c t+\pi K t^2\right)\right], \\ \textstyle
 s_{R X}(t)&=\textstyle \alpha s_{T X}[t-2 R(t) / c],
\end{aligned}
\end{equation}
where \( f_c \) and \( K \) represent the starting frequency and chirp slope, respectively. \( R(t) \) denotes the time-varying distance between the radar and the object, and \( \alpha \) accounts for the signal attenuation due to path loss. The radar receiver performs frequency mixing between the transmitted signal \( s_{TX}(t) \) and the received signal \( s_{RX}(t) \), followed by low-pass filtering to extract the intermediate frequency (IF) signal:
\begin{equation}
\begin{aligned}\textstyle
 s_{\mathrm{IF}}(t)& \textstyle =\alpha {\exp} \left[j (2 \pi f_{\mathrm{IF}} t+\phi_{\mathrm{IF}}\right)],
\end{aligned}
\end{equation}
with the frequency $f_{\mathrm{IF}}$ and phase $\phi_{\mathrm{IF}}$ in the following
\begin{equation}
\begin{aligned} \textstyle
 f_{\mathrm{IF}}& \textstyle =2 K R(t) / c, \\ \textstyle
 \phi_{\mathrm{IF}} & \textstyle = 2\pi f_cR(t)/c+\pi K R(t)^2/c^2  \approx2\pi f_c R(t)/c.
\end{aligned}\label{eq:fre}
\end{equation}
As indicated in Eq.~\ref{eq:fre}, the linear mapping from distance  \( R(t) \)  to  IF frequency  \( f_{\mathrm{IF}} \)  allows a range FFT to segment the IF signal into range bins, each corresponding to a specific object. The frequency-domain representation of the IF signal is:
\begin{equation}\textstyle
\begin{aligned}\textstyle
S_{IF}(t) = \mathcal{F}\{s_{IF}(t)\} & = \alpha \exp\left[j 4 \pi f_c R(t) / c\right] \\
& = \alpha \exp \left[j 4 \pi f_c (R_0 + d(t)) / c\right].
\label{IF signal}
\end{aligned}
\end{equation}
The second equation in Eq. \ref{IF signal} follows by substituting$R(t) = R_0 + d(t)$, where $R_0$ is the static distance and $d(t)$ captures the sound-induced displacement of the object surface.

\subsection{Static Interference in Vibration Sensing} \label{sec: conventional}
In practice, the captured mmWave signal contains reflections not only from vibrating objects but from static surroundings such as walls and furniture. Consequently, the extracted IF signal  in Eq. \ref{IF signal} should be modeled as a superposition of the vibrating signal $S_{vibrate}$ from the vibrating objects and the static interference  $S_{static}$ from the stationary reflectors, where
\begin{equation}
\begin{aligned}
S_{IF} &= S_{vibrate} + S_{static} \\
       &= \alpha \exp \left[j \frac{4 \pi f_c}{c} (R_0 + d(t)) \right] + S_{static}.
\end{aligned}
\label{IF signal practical}
\end{equation}
Eq.~\ref{IF signal practical} shows that the IF signal traces an arc centered at  $S_{static}$ in the IQ plane. Prior methods \cite{zhang2022ambiear, wang2022mmeve} approximate this arc as a circular segment and shift the estimated center to the origin to eliminate the static component $S_{static}$  for vibration extraction. However, sound waves can also induce subtle vibrations in objects typically considered static. As a result, $S_{static}$  becomes time-varying, which interferes with accurate vibration extraction.

\begin{figure}[h] 
\vspace{-0.1in}
    \centering
    \begin{subfigure}[b]{0.24\textwidth}
    \centering
        \includegraphics[width=0.8\linewidth]{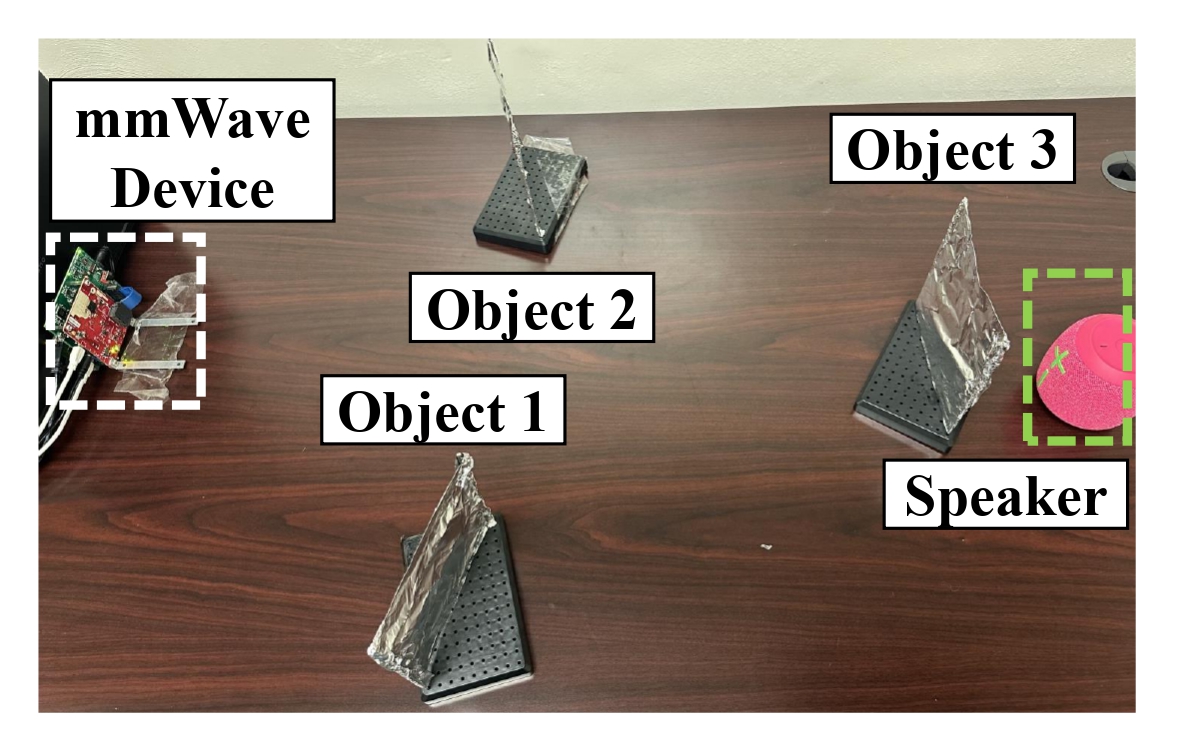}
        \vspace{-0.05in}
        \caption{P1 layout.}
        \label{fig:p1_layout}
    \end{subfigure}
        \begin{subfigure}[b]{0.24\textwidth}
        \centering
        \includegraphics[width=0.8\linewidth]{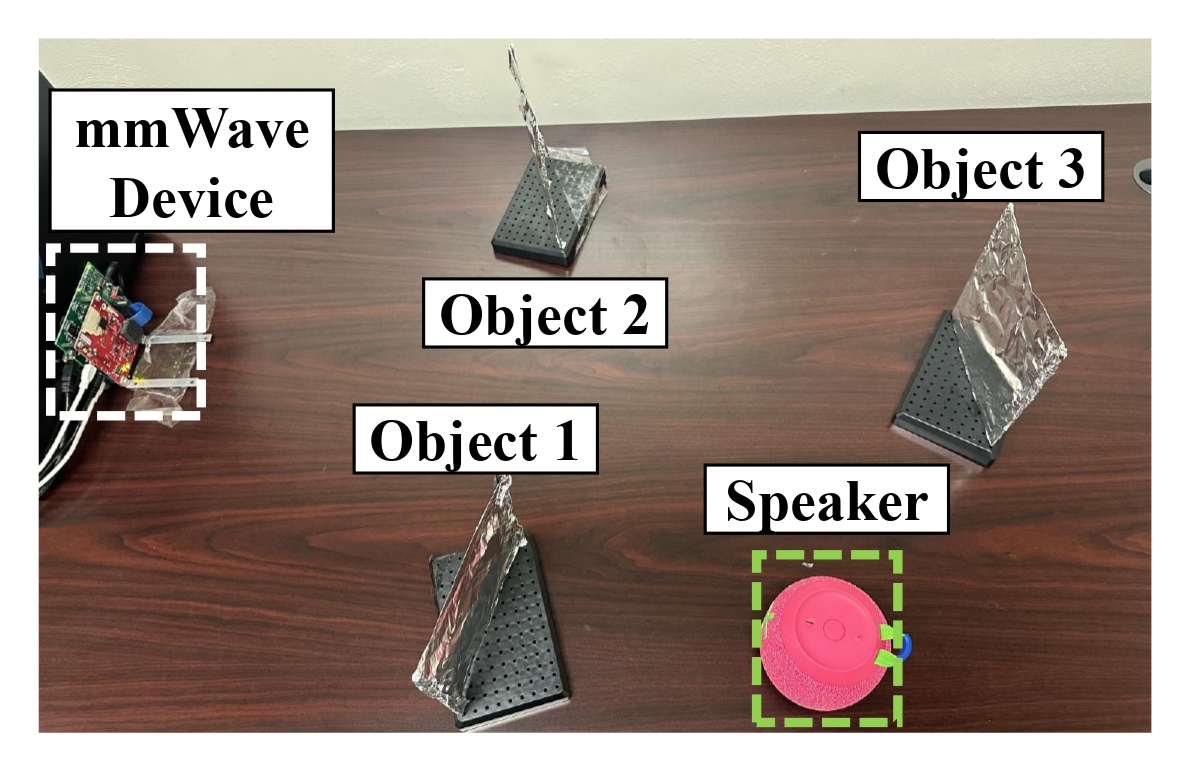}
              \vspace{-0.05in}
        \caption{P2 layout.}
        \label{fig:p2_layout}
    \end{subfigure}
    \begin{subfigure}[b]{0.24\textwidth}
        \centering
        \includegraphics[width=0.8\linewidth]{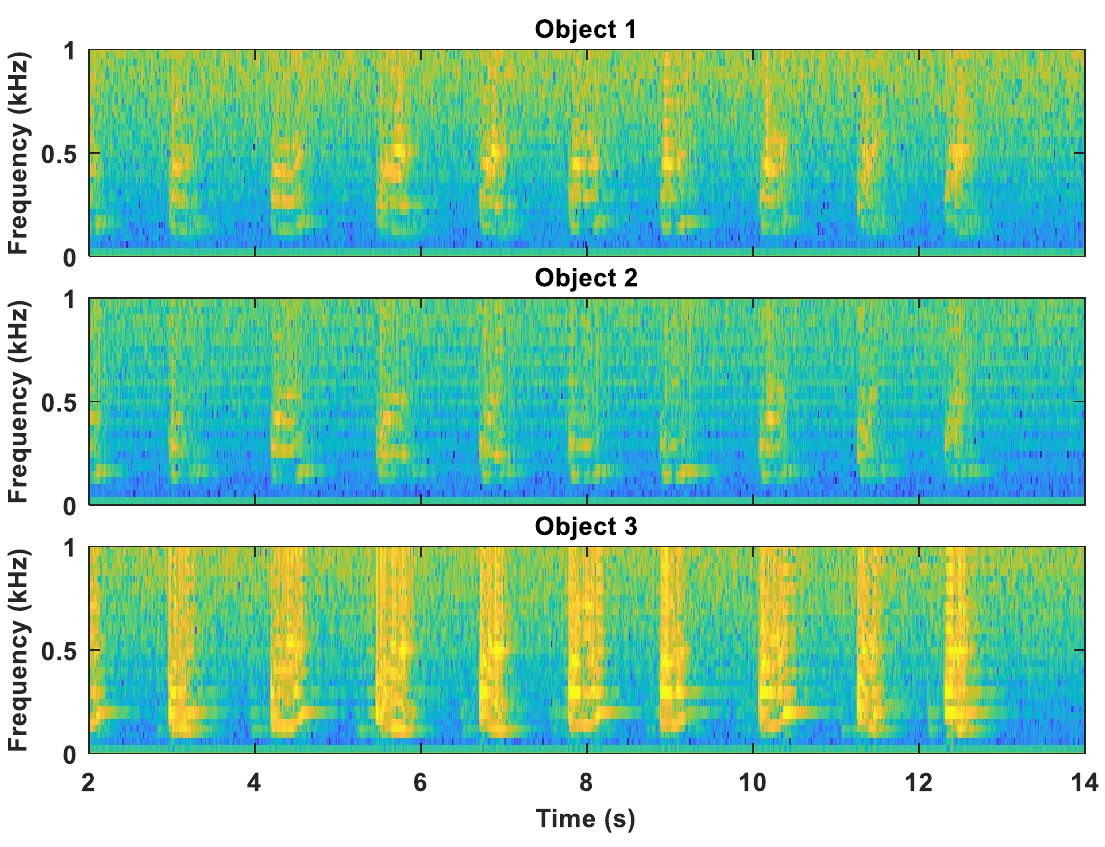}
         \vspace{-0.05in}
        \caption{P1 spectrogram.}
        \label{fig:p1_stft}
    \end{subfigure}
    \begin{subfigure}[b]{0.24\textwidth}
        \centering
        \includegraphics[width=0.8\linewidth]{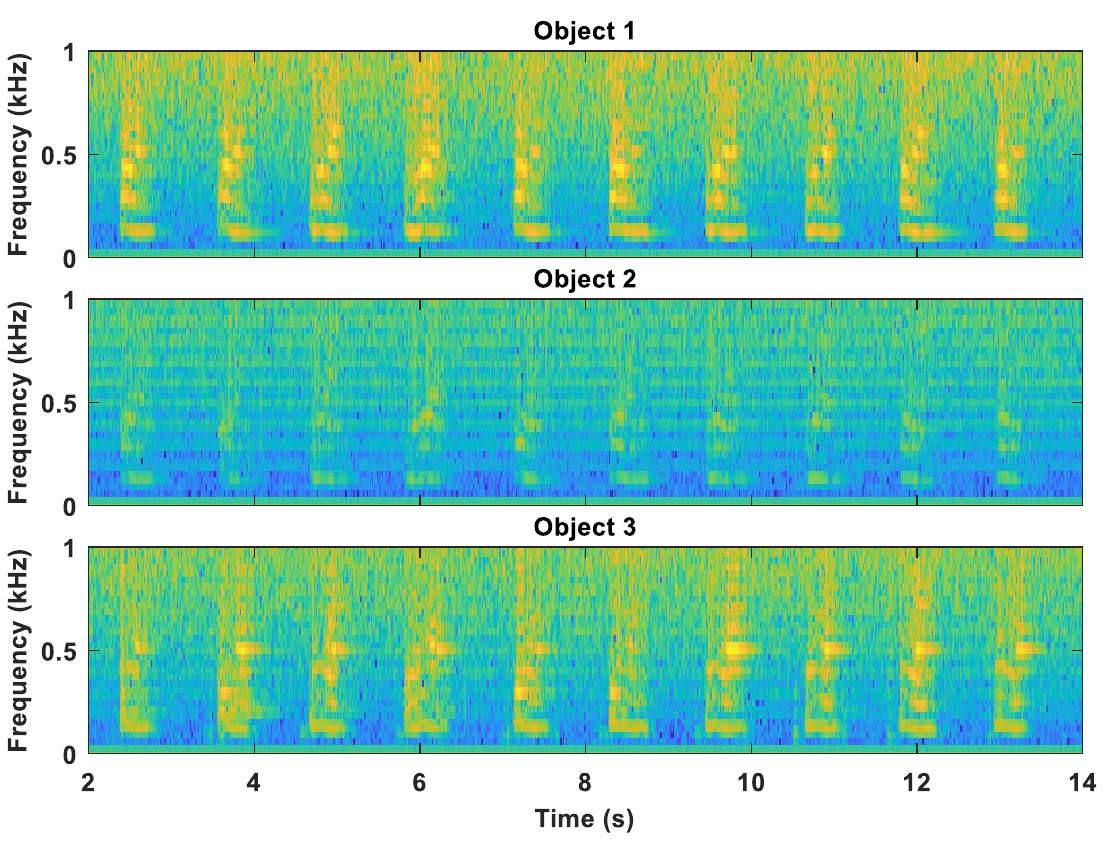}
         \vspace{-0.05in}
        \caption{P2 spectrogram.}
        \label{fig:p2_stft}
    \end{subfigure}
    \vspace{-0.2in}
    \caption{Illustration of frequency responses of speech-induced vibrations on objects with a speaker changing position slightly.}
    \label{fig: vibration_response}
    \vspace{-0.2in}
\end{figure}

\section{EMPIRICAL STUDY}
\subsection{Object Vibration Model Validation} \label{sec: feasibility}
To empirically validate the distance-dependent vibration response from Sec.~\ref{sec:vibration_model}, we conduct an experiment with two layouts in Figs.~\ref{fig:p1_layout} and~\ref{fig:p2_layout}, each containing three fixed tinfoil objects.  A commercial speaker (Wonderboom 3~\cite{a2024_logitech}) is placed at different positions and  relative distances to the objects in each layout.  The speaker plays the spoken digit ``zero'' ten times in each configuration. A 77GHz mmWave radar (TI AWR1843) records the resulting surface vibrations by capturing phase variations in the reflected signals. We extract the vibration signals and apply short-time Fourier transform (STFT) to obtain their spectrograms, shown in Fig. \ref{fig:p1_stft} and~\ref{fig:p2_stft}.

\noindent\textbf{Observation 1:} The comparison of Fig. \ref{fig:p1_stft} and \ref{fig:p2_stft} reveals that the spectrograms of object vibrations vary significantly across different layouts. It results from changes in the relative distances between the speaker and each object: closer objects (e.g., Object 1) exhibit stronger frequency responses, while farther ones (e.g., Object 3) show weaker responses. These findings align with the vibration model in Sec.~\ref{sec:vibration_model}, which indicates that vibration amplitude and frequency content depend on an object’s proximity to the sound source. \textit{This observation motivates the use of spatial differences in frequency responses as a discriminative feature for participant identification.}

\noindent\textbf{Observation 2:}
Within each layout, although the same speech content is played, the vibration spectrograms of different objects exhibit distinct patterns in Fig.~\ref{fig:p2_stft}. These differences arise from the objects’ varying positions, which lead to different captured speech components as well as unique noise and attenuation characteristics.
\textit{This observation suggests that combining  speech-related information across multiple vibration signals yields a more accurate original speech reconstruction.} 

\begin{figure}[H]
\vspace{-0.1in}
    \centering
    \begin{subfigure}[b]{0.47\linewidth}
        \centering
        \includegraphics[width=0.9\linewidth]{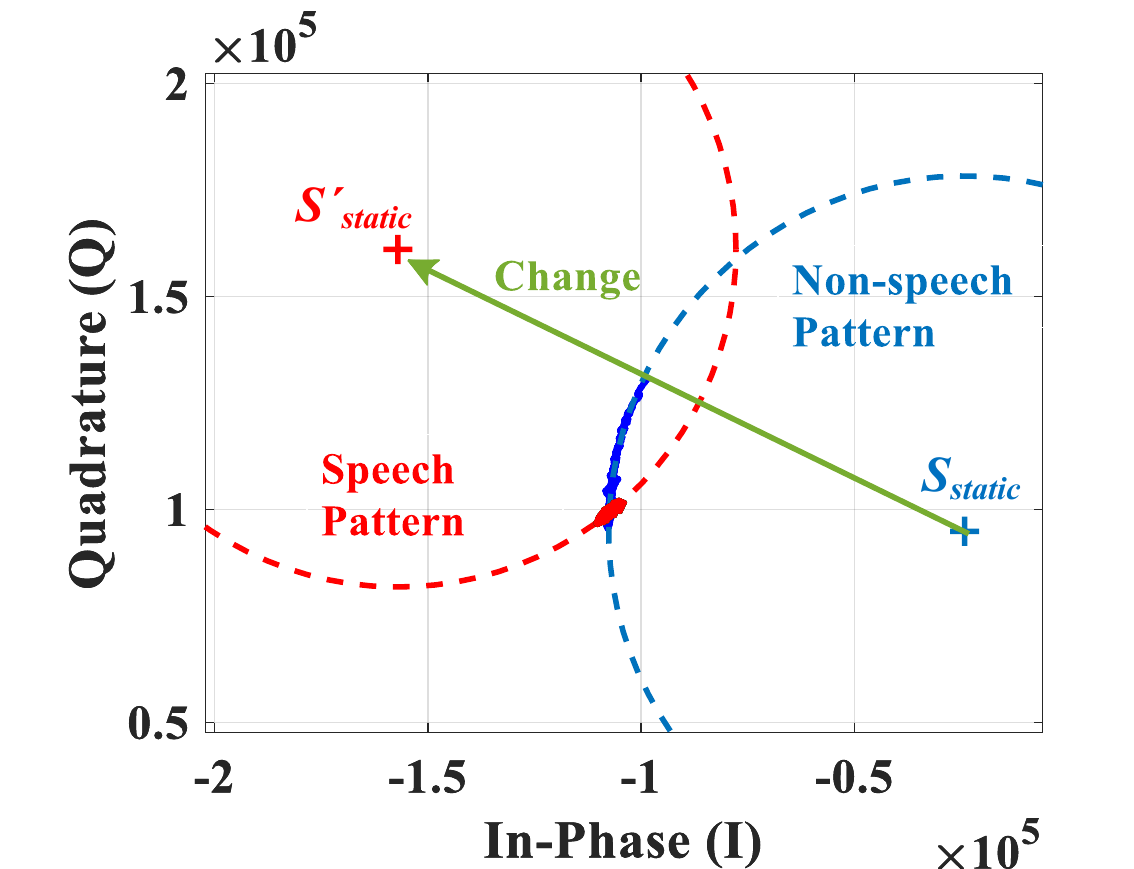}
        \vspace{-0.05in}
        \caption{Different signal patterns.}
        \label{fig:Speech_Non_speech}
    \end{subfigure}
    \begin{subfigure}[b]{0.48\linewidth}
        \centering
        \includegraphics[width=0.85\linewidth]{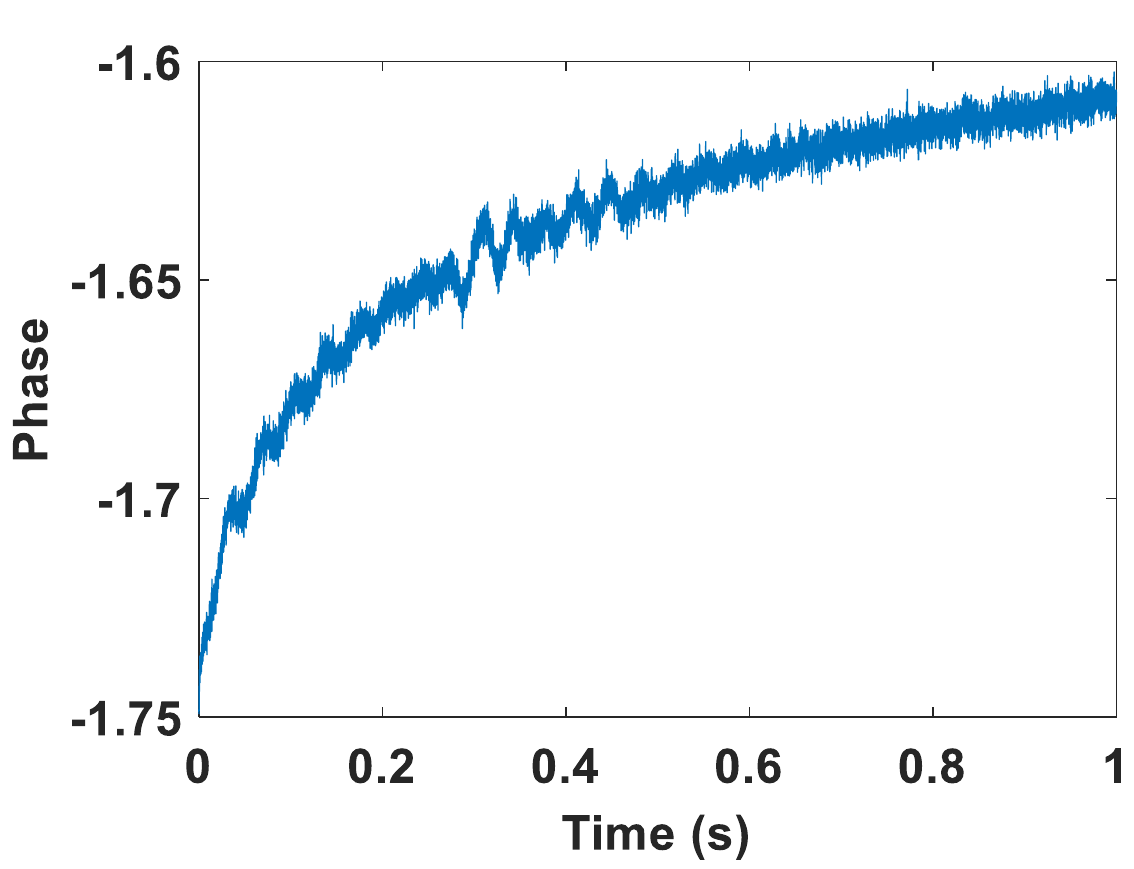}
        \vspace{-0.05in}
        \caption{Silent periods phase changes.}
        \label{fig:phase_change}
    \end{subfigure}
    % \vspace{-0.1in}
    \caption{Static interference Verification.}
    \vspace{-0.2in}
    \label{fig:signal_distortion}
\end{figure}

\begin{figure*}[t]
\centerline{\includegraphics[width=0.85\linewidth]{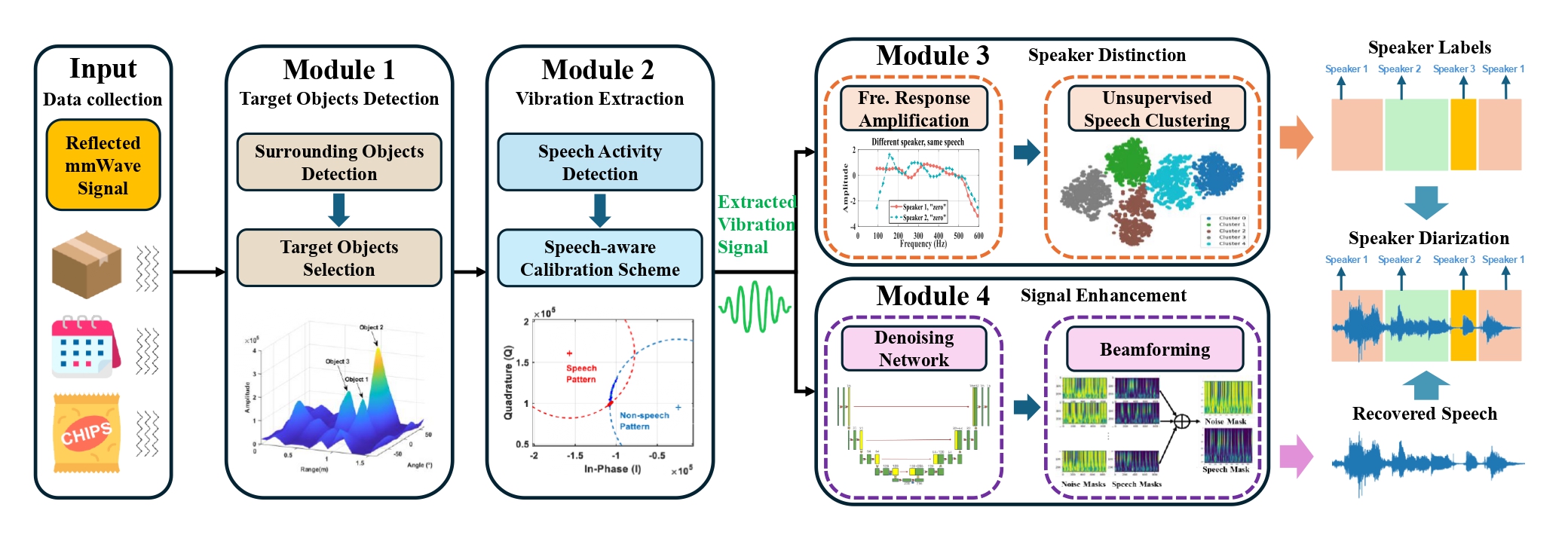}}
\vspace{-0.1in}
\caption{System overview.}
\vspace{-0.25in}
\label{fig:overview}
\end{figure*}

\subsection{Static Interference Verification} \label{sec:phase_distortion}
To examine static interference described in Sec. \ref{sec: conventional}, we visualize the reflected signal from Object 3 in P2 layout in the IQ plane.  Fig.\ref{fig:Speech_Non_speech} reveals that the signal trajectories differ significantly 
depending on whether speech is present. Specifically, the trajectory during silent periods (blue) in Fig. \ref{fig:phase_change} reflects hardware-induced phase changes\cite{jiang2021sense}, while the trajectory during speech (red) exhibits a shifted center, suggesting an altered reflection path. This shift arises because sound waves subtly change the physical state of surrounding objects, leading to altered static interference in the reflected signal. To capture this behavior, the IF signal in Eq.~\ref{IF signal practical} should be revised as:
\begin{equation}\textstyle
S_{IF}^{abs} =  \alpha \exp \left[j 4 \pi f_c R_0 / c\right] + S_{static}^{'},
\label{eq:speech_absence}
\end{equation}
where $S_{static}^{'}$  denotes the altered static interference. From Eq. \ref{eq:speech_absence}, the trajectory center on IQ plane can shift between speech and non-speech segments. Hence, conventional circle-fitting methods \cite{wang2022mmeve, zhang2022ambiear}  inaccurately estimate the center, causing failure to isolate the vibration signal from static interference.

\section{THREAT MODEL}
We consider a threat model in which a confidential meeting is held inside a soundproof room furnished with everyday objects such as calendars, paper bags, and cardboard boxes. Multiple participants are seated around a table, engaging in sensitive conversations. As each participant takes turns speaking, the emitted acoustic waves induce minute surface vibrations on nearby objects.

An external attacker, positioned outside the room, seeks to exploit these vibrations to reconstruct the spoken content. To this end, the attacker deploys a mmWave radar without any physical access to the room. This enables a fully passive and undetectable attack, as it does not rely on in-room devices such as microphones. Furthermore, by leveraging the penetration capability of mmWave signals, the adversary can capture speech-induced vibrations through soundproofing materials, enabling a through-wall attack. More importantly, the attack does not require prior knowledge of the room layout, participant locations, or conversation content, making it practical and applicable to a wide range of real-world environments.

\section{System Design}
An overview of the proposed attack system is illustrated in Fig.~\ref{fig:overview}.  With the collected mmWave data, \textit{Module 1}  selects reflections from objects exhibiting stronger speech-induced vibrations as target signals. \textit{Module 2} mitigates interference caused by nearby objects for accurate extraction of vibration signals. The resulting spectrograms are then processed in parallel. Particularly, \textit{Module 3} amplifies weak vibration frequency components by suppressing self-noise and residual noise to enable reliable speech distinction.  \textit{Module 4} aggregates signals across multiple objects for speech signal enhancement. Finally, the outputs of Modules 3 and 4 are fused to infer both the speech activity with speaker labels and to reconstruct an intelligible speech signal for passive eavesdropping. The following subsections detail each module.

\subsection{Module 1: Target Objects Detection}
\subsubsection{Surrounding Objects Isolation}
\begin{figure}[H]
 \vspace{-0.05in}
\centerline{\includegraphics[width=0.95\linewidth]{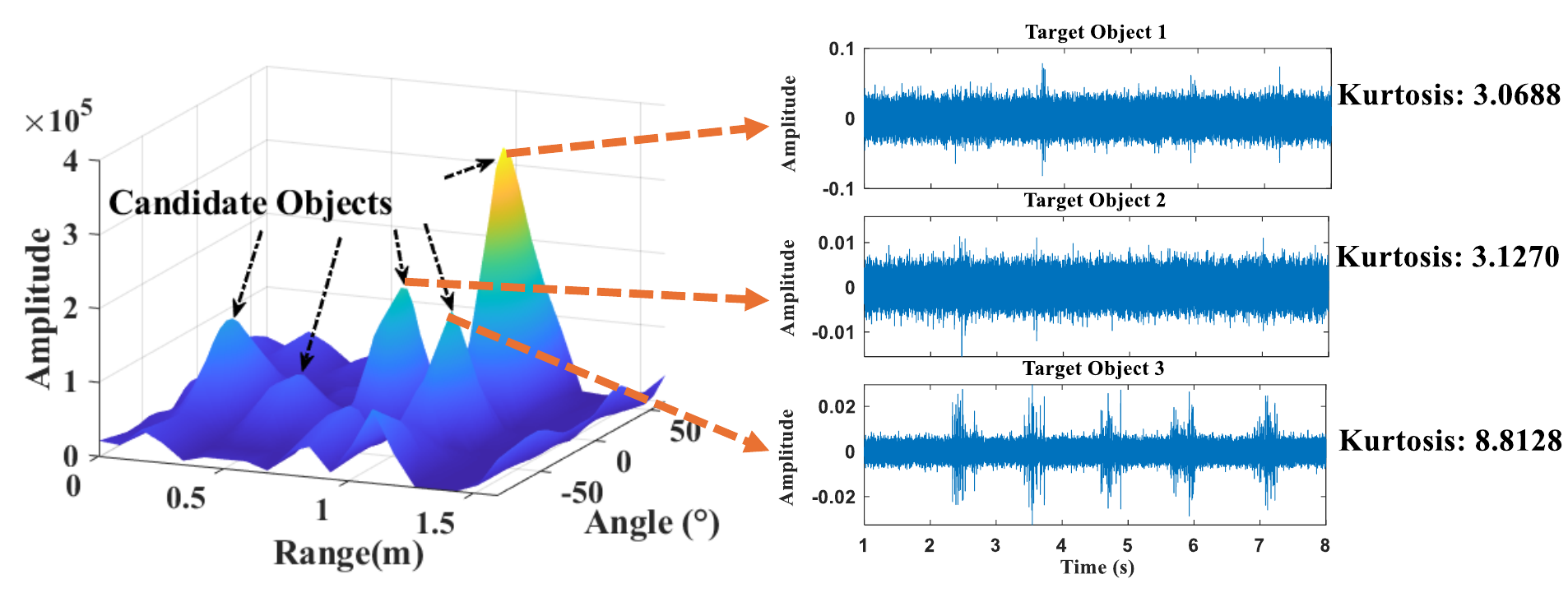}}

\caption{
Object distance/angle identification and target objects selection with the strongest vibration responses}
\vspace{-0.1in}
\label{fig:range_angle}
\end{figure}

The received mmWave signal includes reflections from diverse objects, ranging from highly responsive items (e.g., chip bags) to static structures (e.g., walls). To isolate potential vibration sources, we apply \textit{Range FFT}~\cite{iovescu2017fundamentals} to segment reflections by distance into range bins.
However, objects at the same distance but different directions are indistinguishable in range alone. To address this, we estimate the angle of arrival (AoA) using signals from multiple receiver antennas, resulting in a 2D range-angle map as shown in Fig.~\ref{fig:range_angle}. We then identify candidate objects by selecting high-amplitude regions from this map, corresponding to surfaces with stronger vibration responses.

\subsubsection{Target Object Selection}
As shown in Eq. \ref{IF signal practical}, larger surface displacements result in greater phase variations in the reflected signals. We identify candidate objects by analyzing their phase trajectories for significant dynamics, which appear as abrupt transitions in Fig. \ref{fig:range_angle}. To quantify such variations, we compute the kurtosis of the phase distribution~\cite{shi2023privacy, westfall2014kurtosis}, a statistical measure of tail extremity. The top M objects by kurtosis values are selected as targets for subsequent analysis.

\begin{figure*}[t]
    \centering
    % --- Left: single figure ---
    \begin{minipage}[b]{0.35\linewidth}
        \centering
        \includegraphics[width=0.9\linewidth]{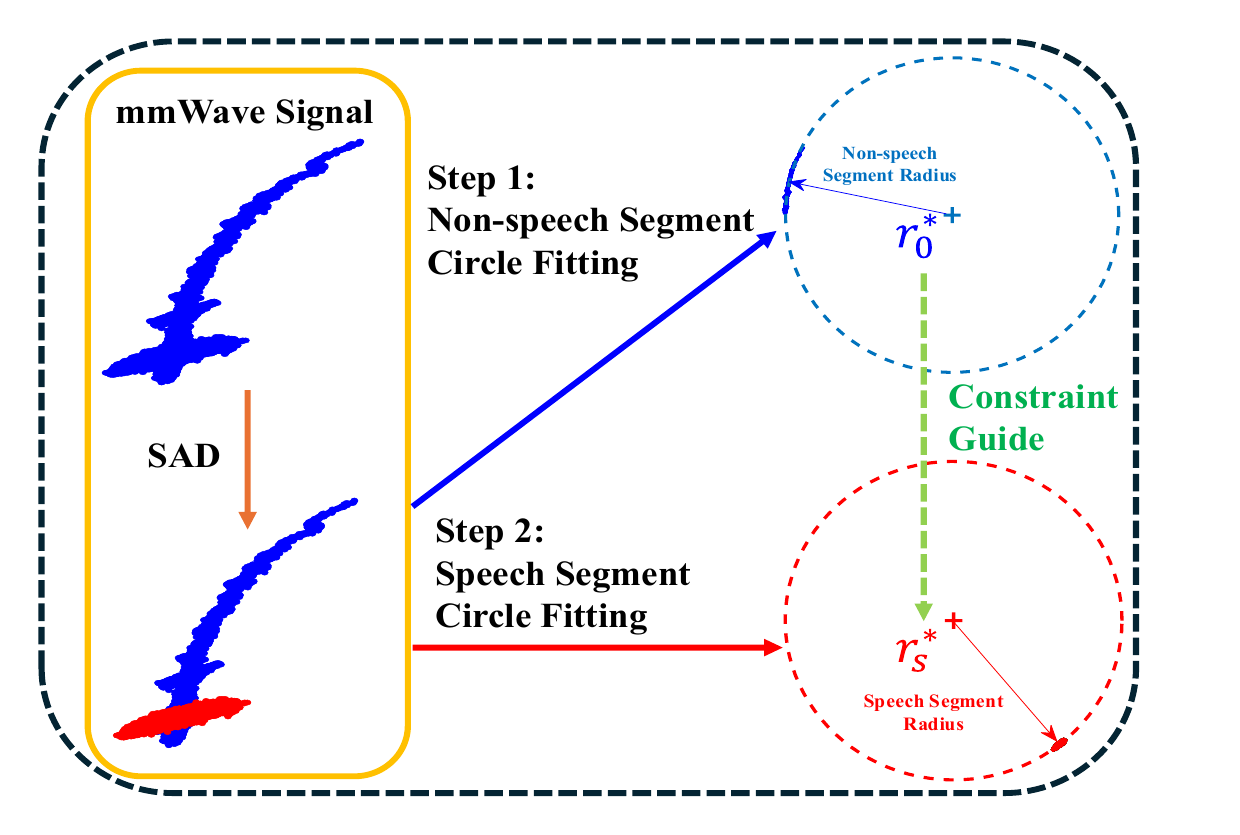}
        \caption{Speech-aware calibration scheme.}
        \label{fig:speech_aware}
    \end{minipage}
    \hfill
    % --- Right: four subfigures ---
    \begin{minipage}[b]{0.63\linewidth}
        \centering
        \begin{subfigure}[b]{0.24\linewidth}
            \centering
            \includegraphics[width=\linewidth]{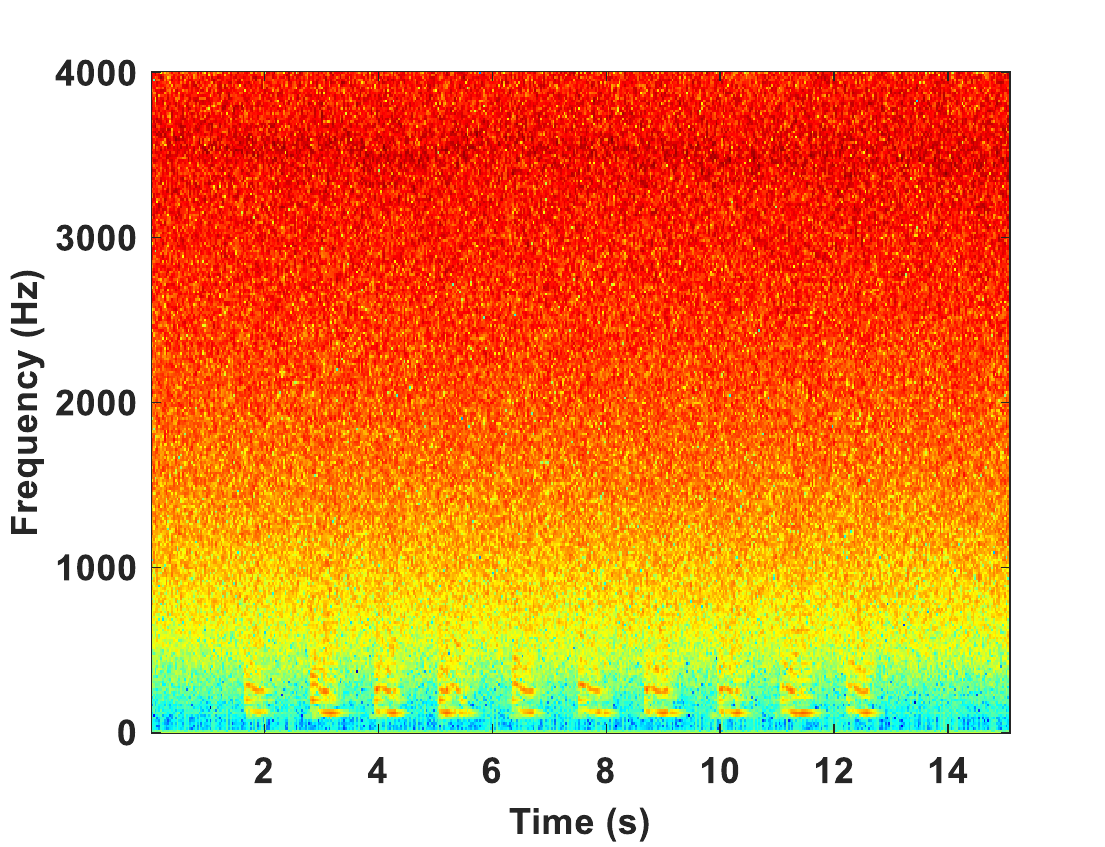}
            \caption{Original spectrogram.}
            \label{fig:process_ori}
        \end{subfigure}
        \hfill
        \begin{subfigure}[b]{0.24\linewidth}
            \centering
            \includegraphics[width=\linewidth]{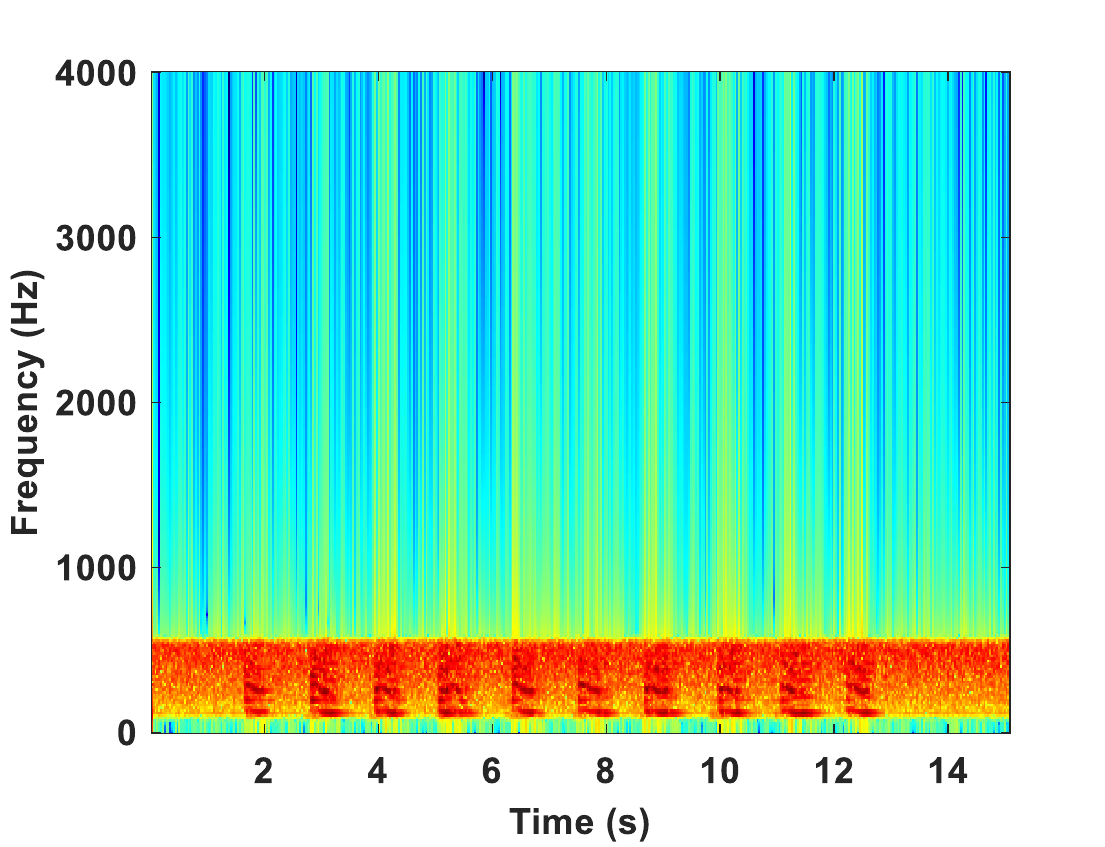}
            \caption{Speech reservation.}
            \label{fig:process_band}
        \end{subfigure}
        \hfill
        \begin{subfigure}[b]{0.24\linewidth}
            \centering
            \includegraphics[width=\linewidth]{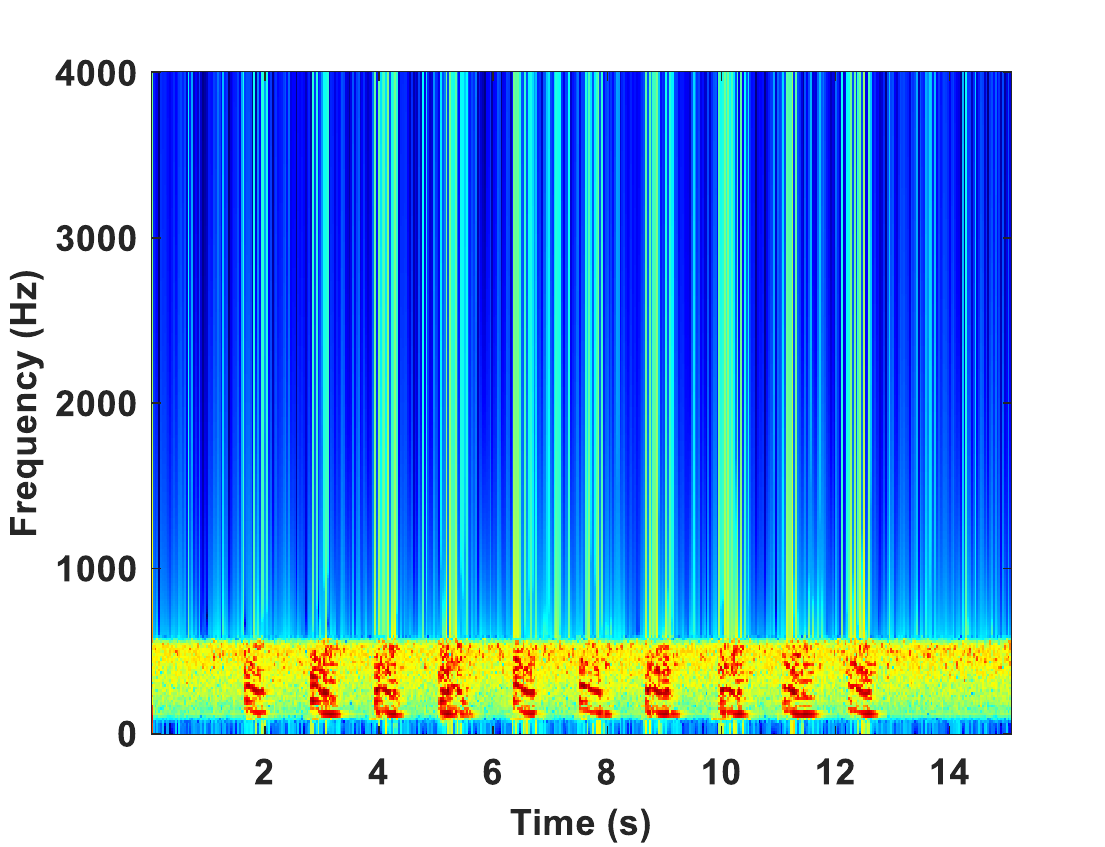}
            \caption{Self-noise attenuation.}
            \label{fig:process_sub}
        \end{subfigure}
        \hfill
        \begin{subfigure}[b]{0.24\linewidth}
            \centering
            \includegraphics[width=\linewidth]{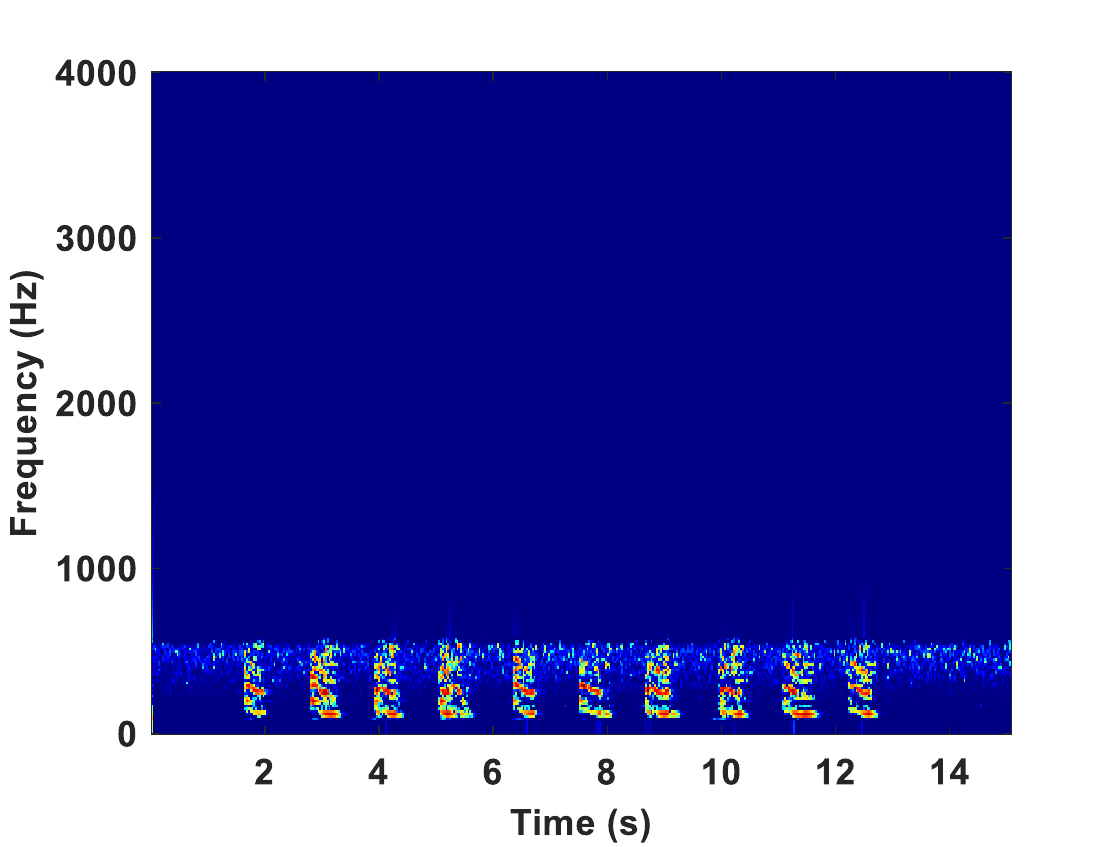}
            \caption{Residuals suppression.}
            \label{fig:process_pow}
        \end{subfigure}
        \vspace{0.05in}
        \caption{(a) Original mmWave spectrogram; (b–d) Spectrogram visualization of mmWave signals at each stage of processing.}
        \label{fig:Res_speaker_layout}
    \end{minipage}

    \vspace{-0.2in}
\end{figure*}

\begin{figure}[H]
\vspace{-0.10in}
    \centering
    \begin{minipage}[b]{0.48\linewidth}
        \centering
        \includegraphics[width=\linewidth]{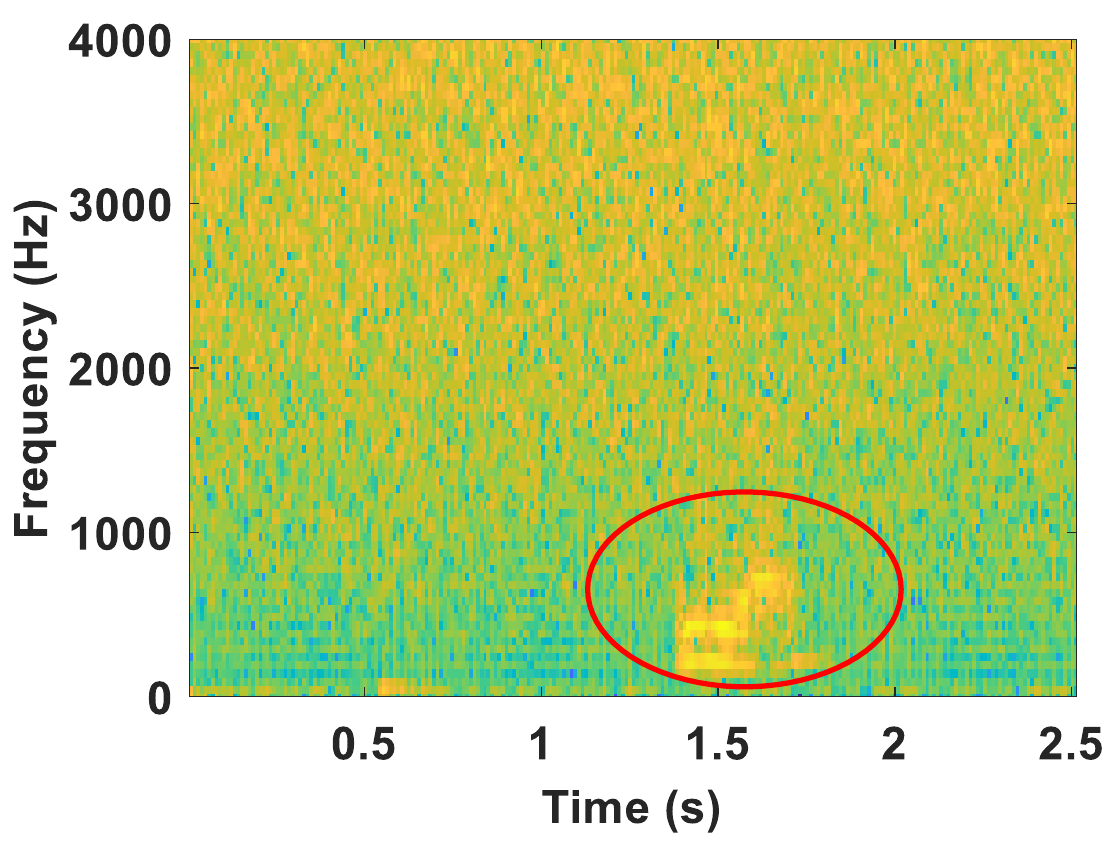}
        \vspace{-0.2in}
        \caption{Signal spectrogram.}
        \label{fig:STFT_before}
    \end{minipage}
    \hfill
    \begin{minipage}[b]{0.48\linewidth}
        \centering
        \includegraphics[width=\linewidth]{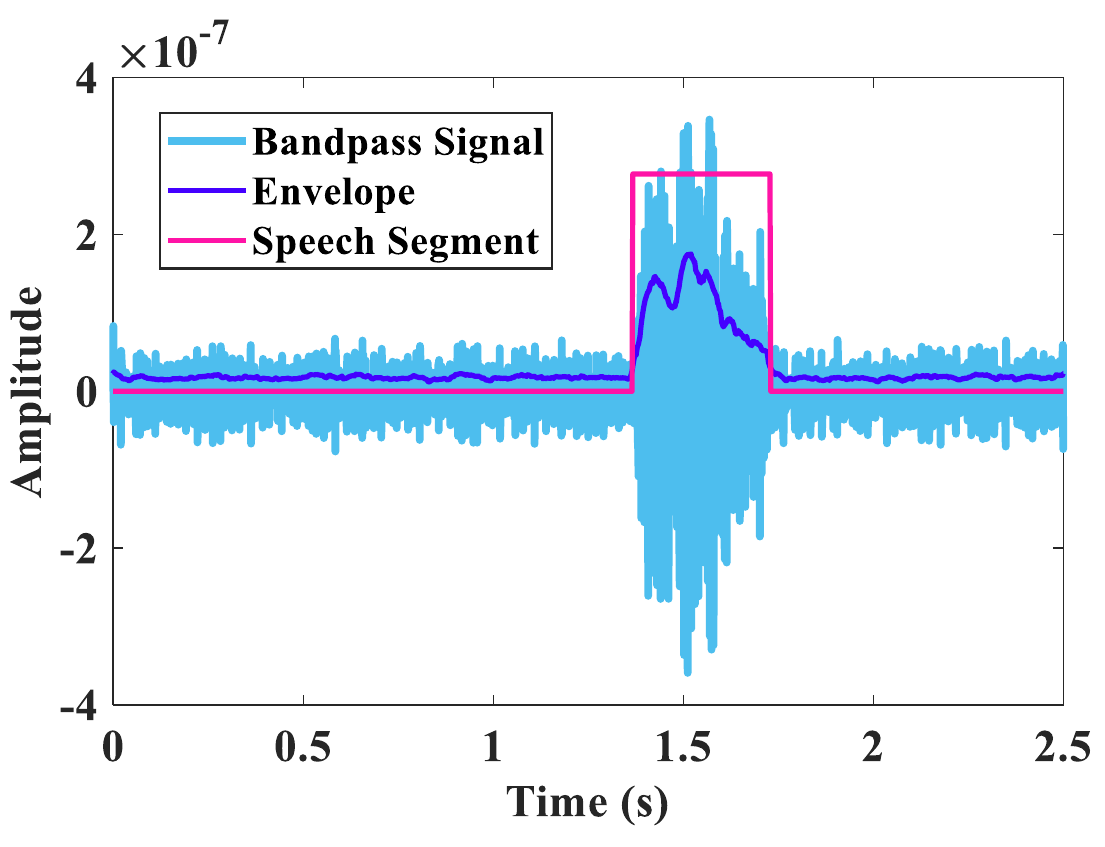}
        \vspace{-0.2in}
        \caption{Bandpass signal.}
        \label{fig:phase_change_after}
    \end{minipage}
\vspace{-0.05in}
    \label{fig:voice_activity_detection}
\end{figure}
\subsection{Module 2: Vibration Extraction}

\subsubsection{Speech Activity Detection (SAD)}
The mmWave radar continuously captures reflected signals from objects in the room, which include both speech and non-speech segments. To isolate speech-related activity, we first analyze the signal spectrogram as in Fig. \ref{fig:STFT_before}, where speech energy is primarily concentrated in the 50–1kHz range, while higher frequencies are dominated by noise. For each object, we apply bandpass filter to retain the speech-relevant frequency band and suppress high-frequency noise. We then calculate the envelope of the filtered signal to detect speech-associated temporal variations. Segments where the envelope exceeds a predefined threshold are identified as speech intervals. Fig. \ref{fig:phase_change_after} shows these segments exhibit clear fluctuations that correlate with speech activity.
\subsubsection{Speech-aware Calibration Scheme} \label{sec:constrained}
Accurately extracting the vibration signal requires removing the static interference, which differs between speech and non-speech periods, as shown in  Eq. \ref{eq:speech_absence}. Despite this difference, both segments share the same signal amplitude $\alpha$.  In addition, Fig. \ref{fig:Speech_Non_speech} shows that the non-speech segment forms a longer arc than the speech segment, providing a more stable basis for estimating the signal circle. Therefore, we use the radius derived from the non-speech segment to accurately determine the circle center for the speech segment, for which we propose a speech-aware calibration scheme. This speech-aware calibration scheme 
is illustrated in Fig.~\ref{fig:speech_aware}.
% \begin{figure}[H]
% \vspace{-0.2in}
% \centerline{\includegraphics[width=0.65\linewidth]{Acc_figures/Speech_aware.pdf}}
%  \vspace{-0.10in}
%  \caption{Speech-aware calibration scheme.}
%  \vspace{-0.10in}
% \label{fig:speech_aware}
% % \vspace{-0.05in}
% \end{figure}
\noindent\textit{Step 1: Non-Speech Segment Circle Fitting.} Let $\mathbf{X}=\left\{x_n\right\}, x_n \in \mathbb{R}^2$ denote the sample points of the non-speech segment. We fit a circle to these points by minimizing the total distance between the samples and the estimated circle, resulting in the center \(c_0\) and radius \(r_0\):
\begin{equation}
\begin{aligned}\label{Eq: non-speech segment radius}
c_0^*, r_0^* = \arg \min\nolimits_{c_0, r_0} \sum\nolimits_{x_n \in \mathbf{X}}\left(\left\|x_n - c_0\right\| - r_0\right)^2. 
\end{aligned}
\end{equation}
\noindent\textit{Step 2: Speech Segment Circle Fitting. } 
The optimal radius \( r_0^* \) obtained from the non-speech segment serves as an initial estimate~\cite{chang2007constrained}, but lacks precision. To refine this estimate,  we use \( r_0^* \) as a constraint to guide the circle fitting for the speech segment. Specifically, we allow the fitted radius to deviate within a margin \(\gamma\), resulting in a feasible range of \([(1-\gamma)r_0^*, (1+\gamma)r_0^*]\).  Let \(\mathbf{Y}=\left\{y_k\right\}, y_k \in \mathbb{R}^2\) represent the IQ samples of the speech segment. The constrained circle fitting problem is then formulated as:
\vspace{-0.05in}
\begin{equation}
\begin{aligned}
c_s^*, r_s^* &=\arg \min\nolimits_{c_s, r_s} \sum\nolimits_{y_k \in \mathbf{Y}} \left( \left\| y_k - c_s \right\| - r_s \right)^2,\\
&\quad \text{s.t. } (1-\gamma)r_0 \leq r_s \leq (1+\gamma)r_0. \label{eq:circlefit}
\end{aligned}
\end{equation}
We employ the Levenberg–Marquardt algorithm~\cite{gavin2019levenberg} to efficiently solve Eq. \ref{eq:circlefit} and obtain the optimal circle center \(c_s^*\). Then we obtain the calibrated phase signal $y(t)=\{y_k-c^*_s\}$ by shifting the IQ coordinate origin for further distinction.

\subsection{Module 3: Speaker Distinction}
Since the extracted vibration frequency response is not pronounced against noises, we first amplify the response with a three-stage signal processing scheme. After that, we employ an unsupervised clustering method for speaker distinction. 

\subsubsection{Frequency Response Amplification} In the amplification, we consider irrelevant high-frequency components, self-noise from hardware imperfection and residual noise in each stage.

\noindent\textit{Stage 1: Speech component reservations.} Speech-induced vibrations mainly reside in low-frequency spectrum~\cite{wang2024vibspeech}. We therefore apply a bandpass filter to the calibrated phase signal \( y(t) \), yielding the filtered signal \( y_{\mathrm{BP}}(t) \).  As illustrated in Fig. \ref{fig:process_ori} and \ref{fig:process_band}, filtered signal preserves all speech-relevant features.

\noindent\textit{Stage 2: Self-noise attenuation.} Self-noise due to imperfections in mmWave hardware remains in the low-frequency band, which is overlapped with the desired speech-induced vibration responses. As the noise is relatively stationary and uncorrelated with the speech content \cite{siddiq2018phase}, we employ the spectral subtraction  to  \( Y(t,f) \), the STFT of \( y_{\mathrm{BP}}(t) \), to the mitigate the noise impact. To begin, we estimate the noise magnitude spectrum from the first \( N \) non-speech frames as $|\hat{D}(f)| = \frac{1}{N} \sum_{t=1}^{N} |Y(t,f)|$. We then calculate the denoised magnitude spectrum with an over-subtraction factor \( \alpha \) and a spectral floor parameter \( \beta \), where we have $|\hat{X}(t,f)| = \max \left( |Y(t,f)| - \alpha \cdot |\hat{D}(f)|,\; \beta \cdot |Y(t,f)| \right)$. In the end, 
% \vspace{-0.05in}
% \begin{equation}
% |\hat{D}(f)| = \frac{1}{N} \sum_{t=1}^{N} |Y(t,f)| \nonumber
% \label{eq:noise_estimate}
% \end{equation}
% Given the estimated noise, we calculate the denoised magnitude spectrum with an over-subtraction factor \( \alpha \) and a spectral floor parameter \( \beta \):
% \begin{equation}
% |\hat{X}(t,f)| = \max \left( |Y(t,f)| - \alpha \cdot |\hat{D}(f)|,\; \beta \cdot |Y(t,f)| \right) \nonumber
% \label{eq:spectral_subtraction}
% \end{equation}
the complex spectrum is constructed by combining the estimated magnitude with the original phase $\angle Y(t,f)$:
\begin{equation}
\hat{X}(t,f) = |\hat{X}(t,f)| \cdot e^{j \angle Y(t,f)}.
\label{eq:phase_restore}
\end{equation}
The spectrogram in Fig.~\ref{fig:process_sub} illustrates the effectiveness of spectral subtraction in mmWave self-noise attenuation.

\noindent\textit{Stage 3: Residuals suppression.} Although stage 2 improves the visibility of speech components, residual noise persists due to imperfect noise estimation~\cite{boll1979suppression}. While typically low in energy, such residuals may still distort weak speech-induced frequency responses. To suppress residual interference, we compute the auto power spectrum of the estimated complex spectrum:
\begin{equation}
\begin{aligned}
P(t,f) &= \hat{X}(t,f) \cdot \hat{X}^*(t,f) \\
       &= |S(t,f)|^2 + |R(t,f)|^2 + 2\Re \{ S(t,f) R^*(t,f) \}. \nonumber
\end{aligned}
\end{equation}
where  \( S(t,f) \) and \( R(t,f) \) indicate the  speech and residual noise. Since \( R(t,f) \) is uncorrelated with the speech signal and has zero mean\cite{siddiq2018phase},  \( \Re \{ S(t,f) R^*(t,f) \} \) averages out across frames. Consequently, the power spectrum primarily reflects the energy of the speech signal with minor noise impacts as:
\begin{equation}
P(t,f) \approx |S(t,f)|^2 + |R(t,f)|^2.  
\label{eq:power_spectrum_approx}
\end{equation}
Since \( |R(t,f)|^2 \) remains small, Eq. \ref{eq:power_spectrum_approx} effectively suppresses residual noise while preserving stable and interpretable speech-induced frequency responses, as shown in Fig.~\ref{fig:process_pow}.

\subsubsection{Unsupervised Speech Clustering}
In practical scenarios, the number of active speakers is unknown to the attacker. To address this, we estimate the number of speaker clusters by computing the Calinski-Harabasz (CH) index across a candidate set \( N \in \{2, 3, \dots, N_{\max}\} \), selecting the value that maximizes the index.
Given the estimated number of speakers, we perform clustering using a Gaussian Mixture Model (GMM)~\cite{reynolds2009gaussian}. The GMM models the underlying distribution of frequency response features and partitions them into coherent clusters, enabling fully unsupervised speaker distinction without requiring any labeled data.

\begin{figure}[H]
    \centering
    \vspace{-0.1in}
    \includegraphics[width=1\linewidth]{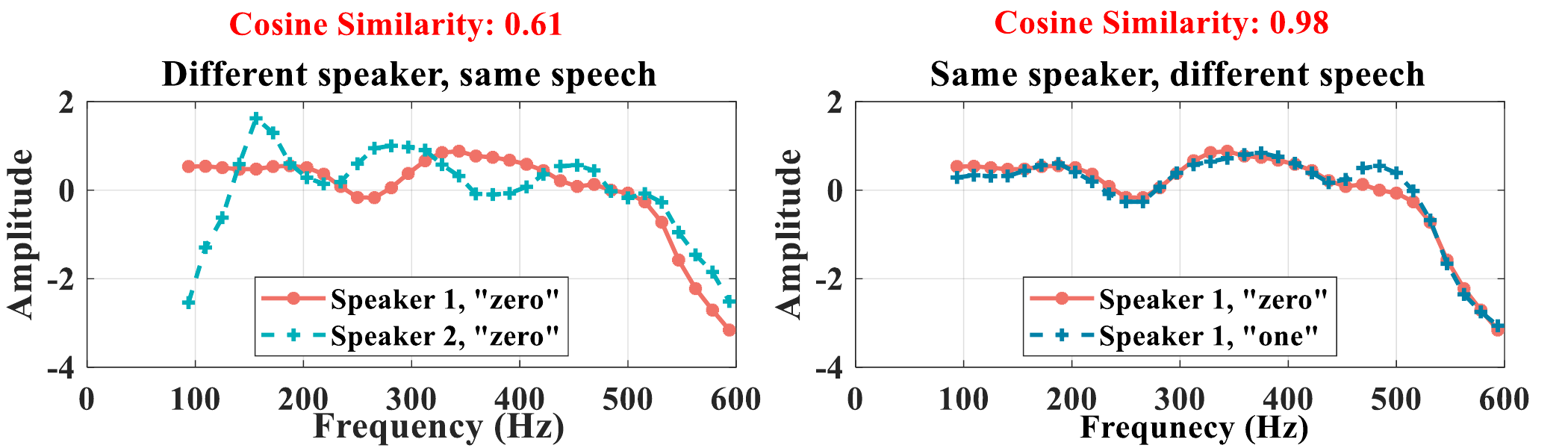}
    \vspace{-0.15in}
    \caption{Spectrum envelope vectors.}
    \label{fig:spectral_sim}
    \vspace{-0.1in}
\end{figure}

However, the high dimensionality (e.g., \(256 \times 256\)) of raw spectrograms $P(t,f)$ hinders efficient clustering. Thus, we extract the spectral envelope $E_m(f)$ from  the spectrogram for each object $m$. The envelope captures the amplitude pattern across frequencies, emphasizing bands with strong speech-induced vibrations and suppressing irrelevant noise. To standardize feature size, each envelope is uniformly sampled to form a low-dimensional vector for downstream distinction. To show that $E_m(f)$ maintains the capability for speaker distinction, we conduct a case study with two speakers at different locations. We use cosine similarities between spectral envelope vectors to measure the capability. As shown in the Fig.~\ref{fig:spectral_sim}, case (i) where different speakers utter the same numbers yields a similarity of 0.61, while case (ii) where the same speaker utters different numbers results in a similarity of 0.98. It proves that the spectral envelope vectors reliably encode speaker-specific spatial characteristics and remain robust to content variations.

To leverage the spatial diversity across multiple objects, we concatenate the spectral envelope vectors from all M selected targets into a unified feature vector for each speech segment as $E(f)=\{E_m(f)\}_M$. This joint representation captures object-specific responses and enhances class separability.

\subsection{Module 4: Signal Enhancement}
\subsubsection{Denoising Network}
To combine vibration signals from multiple objects, we employ Minimum Variance Distortionless Response (MVDR) beamforming in the time-frequency domain without requiring explicit spatial priors such as direction-of-arrival. However, the effectiveness of MVDR depends heavily on accurate steering vector estimation, which is challenged by substantial hardware-induced self-noise in mmWave signals. Although the stage 2 of module 3 applies frequency-domain self-noise attenuation for response amplification, this single-domain approach fails to address the joint time–frequency characteristics on which MVDR relies.

To tackle this limitation,  we adopt an encoder-decoder architecture to remove the inherent noise at time–frequency domain. It consists of $4$ convolutional layers for downsampling and $4$ deconvolutional layers for upsampling. Additionally, $4$ skip connection layers are used to connect the convolutional-deconvolutional layer pairs, preserving important features during the reconstruction process.  The denoising network is trained using the Huber loss function, which provides a balance between L1 and L2 losses and improves robustness to outliers. The network is optimized with the Adam optimizer. Trained on paired noisy and clean spectrograms, the model learns to remove spectral noise while maintaining the integrity of speech-induced frequency components. 

\subsubsection{Beamforming}
After denoising the spectrograms, we estimate the steering vector using a neural-based full-band mask estimation approach. This method jointly infers time-frequency masks across the entire spectrum, effectively capturing the harmonic structure of speech and mitigating the influence of noise components.

\begin{figure}[h]
\vspace{-0.1in}
\centerline{\includegraphics[width=0.95\linewidth]{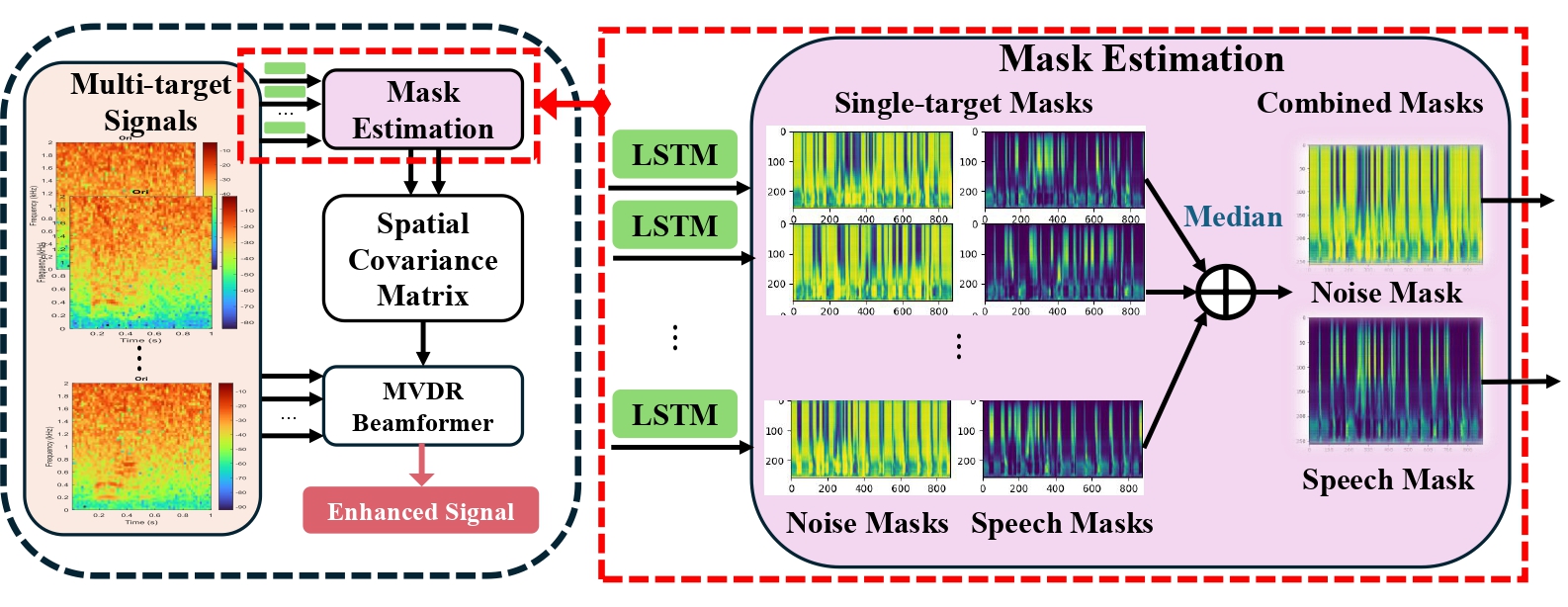}}
\vspace{-0.05in}
\caption{Illustration of using neural-based mask estimation method to combine signals from multiple objects.}
\label{fig:speech_enhancement}
\vspace{-0.1in}
\end{figure}

 Our MVDR beamforming method is illustrated in Fig. \ref{fig:speech_enhancement}. We first generate masks for both speech and noise, which are then used to compute the corresponding spatial covariance matrices. These matrices guide the beamformer to enhance the desired speech signal while suppressing background interference.
To model temporal dependencies in the spectrogram, we employ an LSTM-based network for mask prediction. Our LSTM network is trained with the magnitude spectrum approximation (MSA) loss on clean speech and mmWave noise data. During inference, the network outputs one mask per input signal. Since MVDR requires a single spatial covariance matrix, we aggregate the individual masks across signals using median combination~\cite{erdogan2016improved}, which provides robustness to outliers and yields a representative mask for beamforming.

\section{Evaluation}
\subsection{Experiment Setup}
\begin{figure}[H]
\vspace{-0.2in}
    \centering
    \includegraphics[width=0.7\linewidth]{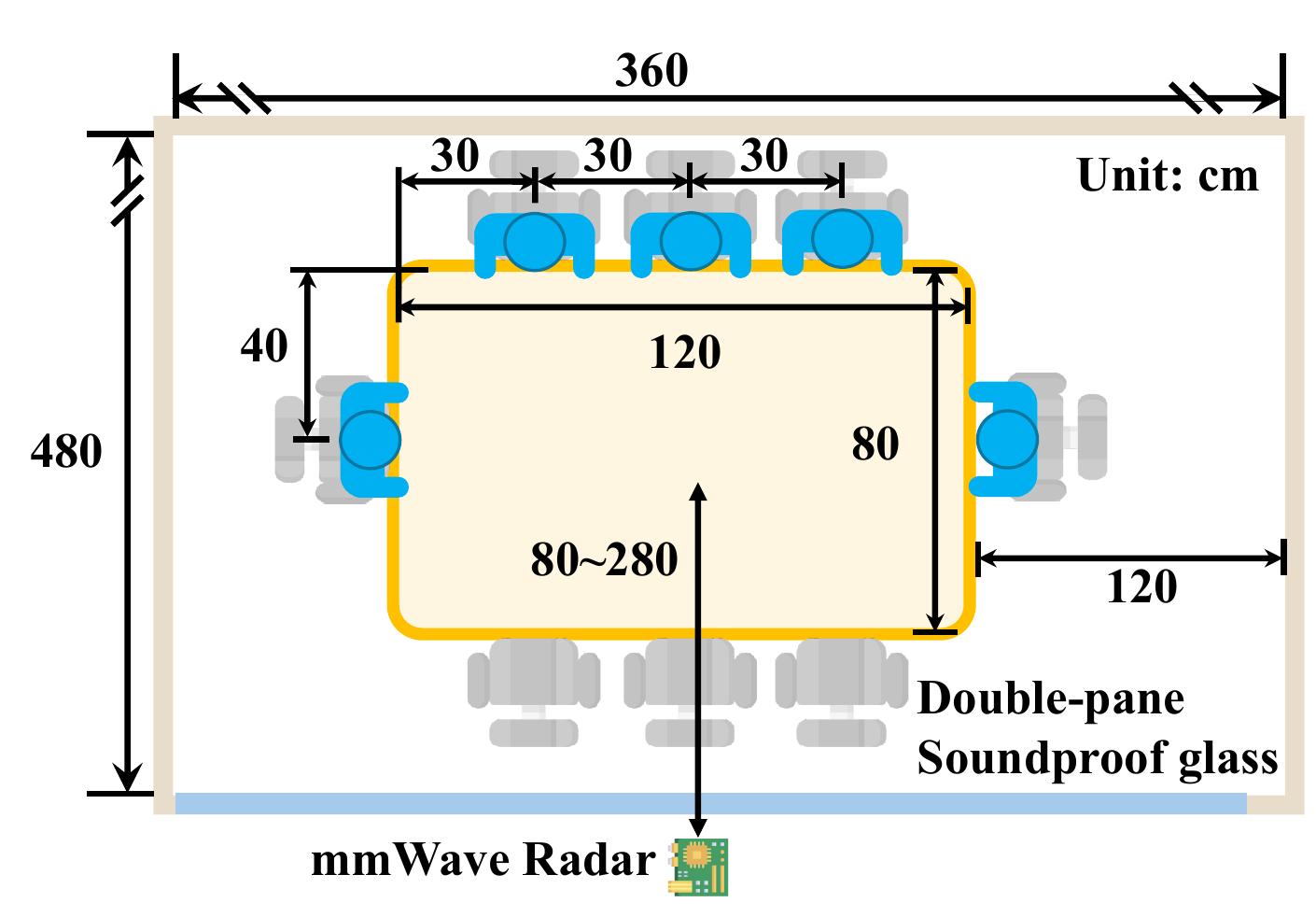}
    \vspace{-0.1in}
    \caption{Room layout.}
    \label{fig:overall_setup}
    \vspace{-0.15in}
\end{figure}

We implement the proposed attack in a soundproof room separated by a double-pane glass window, as shown in Fig. \ref{fig:overall_setup}. A TI AWR1843 mmWave radar (3 TX / 4 RX) and a DCA1000EVM data capture board are placed outside the enclosure, enabling signal acquisition without physical intrusion. We configure the radar to transmit chirp signals in the 77–81GHz band, with each frame consisting of 128 chirps over 10.64ms. The resulting sampling rate is 12.5kHz. The denoising and neural mask estimation networks are implemented in TensorFlow and trained on an NVIDIA RTX 3070 GPU. The tolerance parameter $\gamma$ in the constrained circle fitting method (Sec.~\ref{sec:constrained}) is empirically set to 0.1.
\noindent\textbf{Default Setup}: We use a Wonderboom 3 speaker placed at five different positions to simulate multi-speaker scenarios. Three tinfoil sheets are used as reflective objects and placed on a table at distances of 1.1\,$\mathrm{m}$
, 1.3\,$\mathrm{m}$, and 1.1\,$\mathrm{m}$ from the radar, with angular positions of $0^\circ$, $30^\circ$, and $-30^\circ$, respectively. The room layout is in Fig.~\ref{fig:overall_setup}.

\noindent \textbf{Speech Dataset.} We evaluate our attack using the AudioMNIST dataset~\cite{audiomnist2023}, which includes 25,000 recordings of spoken digits (0–9). To train our models, we synthesize radar-like speech data by mixing clean audio samples from the LibriSpeech corpus~\cite{panayotov2015librispeech} with mmWave recordings collected in the absence of speech. This mixing process emulates speech-induced reflections as captured by mmWave radar. In total, we generate 40,000 synthetic samples and transform them into spectrograms of size 128×128 for training.

We evaluate the propsoed \textit{Speaker Distinction} and \textit{Signal Enhancement} schemes. For speaker distinction, we use the success rate as the  metric, defined as the proportion of correctly classified samples.
For signal enhancement, we report Signal-to-Noise Ratio (SNR) and Peak SNR (PSNR). SNR measures the ratio between the power of the recovered speech and the background noise, reflecting overall enhancement quality. PSNR captures the peak error between the enhanced and clean reference signals, emphasizing fidelity at the most perceptually sensitive regions. We compare our approach (Ours) with  prior work \cite{shi2023privacy, hu2023mmecho}, which recovers speech using only the object with the strongest mmWave vibration response. For fair comparison, we apply the same signal processing pipeline to the strongest vibration signal. We refer to this baseline as \textit{VibStrong}.

\subsection{Experimental Analysis}

\subsubsection{Performance of Speaker Distinction}
We investigate how practical factors such as speaker arrangement, number of objects, object layout, and sensing distance affect the performance of our speaker distinction method.

\begin{figure}[H]
    \centering
    \begin{minipage}[b]{0.42\linewidth}
     \vspace{-0.2in}
        \centering
        \includegraphics[width=0.9\linewidth]{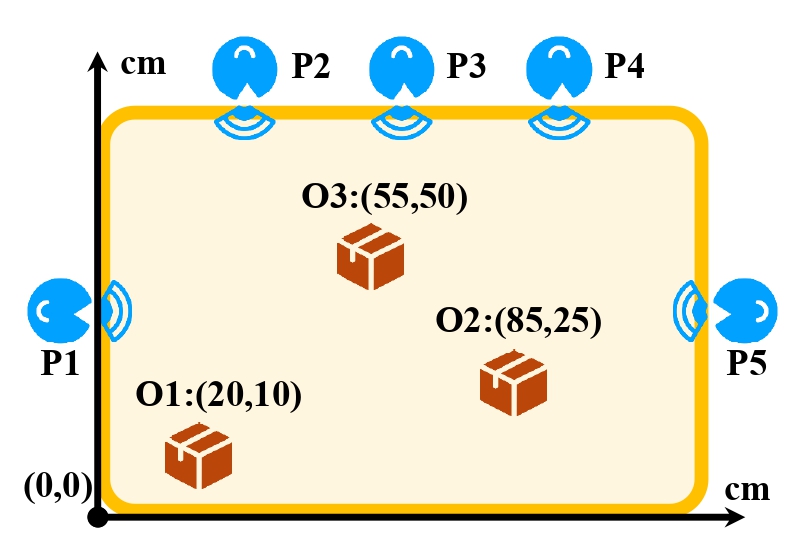}
        \vspace{-0.1in}
        \caption{Layout.}
        \label{fig:SC_setup}
    \end{minipage}
    \hfill
    \begin{minipage}[b]{0.53\linewidth}
     \vspace{-0.2in}
        \centering
        \includegraphics[width=\linewidth]{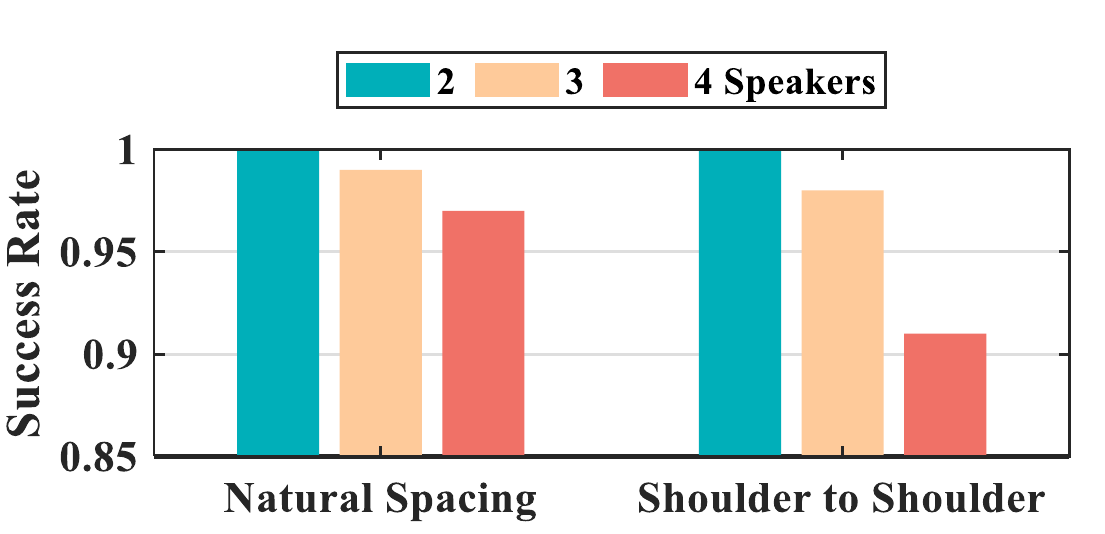}
         \vspace{-0.2in}
        \caption{Arrangement impact.}
        \label{fig:speaker_acc}
    \end{minipage}
    % \caption{(a) Experimental layout. (b) Classification performance across different speaker arrangements.}
    \label{fig:speaker_layout_combo}
    % \vspace{-0.2in}
\end{figure}

\noindent \textbf{Impact of Speaker Arrangement.} We evaluate how the seating arrangement of speakers affects the distinction performance by comparing two types of layouts: \textit{Shoulder to Shoulder} where speakers are in close proximity, and \textit{Natural Spacing} where they are more spatially separated, as illustrated in Fig. \ref{fig:SC_setup} and detailed in Table \ref{tab:layout_setting}. The results in  Fig. \ref{fig:speaker_acc} show that \textit{Natural Spacing} achieves a superior success rate due to its more distinguishable vibration pattern. Nonetheless, the attack remains robust under the challenging \textit{Shoulder to Shoulder} layout, achieving success rate above 0.9 with 4 speakers.
\begin{table}[H]
    \centering
    \vspace{-0.2in}
    \begin{tabular}{l|ccc}
        \specialrule{1.5pt}{1pt}{1pt}
        \textbf{Arrangement/\# of speakers} & 2 & 3 & 4 \\
        \specialrule{1pt}{1pt}{1pt}
        \textbf{Natural spacing} & P1 P5 & P1 P3 P5 & P1 P3 P4 P5 \\
        \textbf{Shoulder to shoulder} & P4 P5 & P3 P4 P5 & P1 P2 P3 P4 \\
        \specialrule{1.5pt}{1pt}{1pt}
    \end{tabular}
    \caption{Speaker positions under different arrangements.}
    \label{tab:layout_setting}
    \vspace{-0.2in}
\end{table}

\begin{figure}[H]
 \vspace{-0.1in}
    \centering
    \begin{minipage}[b]{0.47\linewidth}
        \centering
        \includegraphics[width=\linewidth]{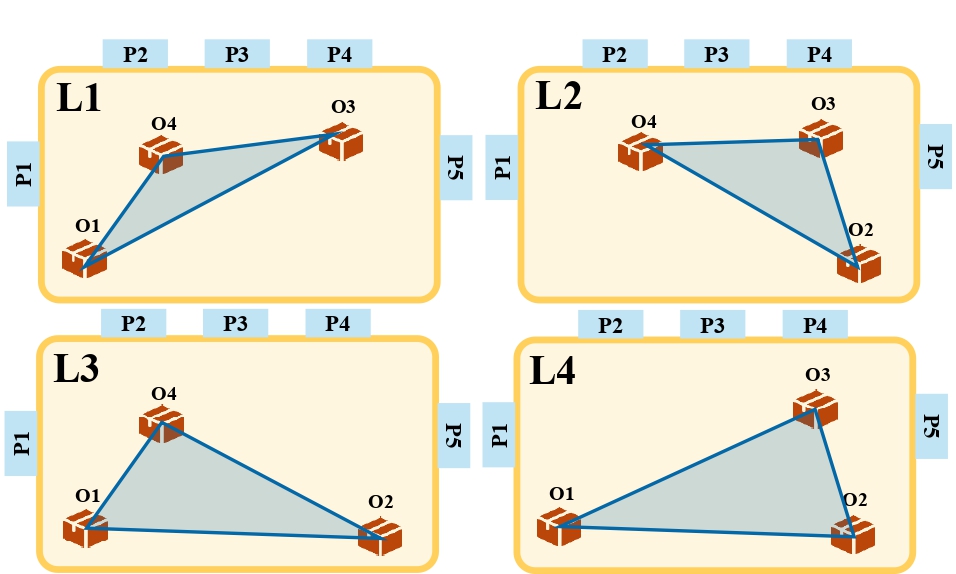}
         \vspace{-0.2in}
        \caption{3-object layouts.}
        \label{fig:object_layout_setup}
    \end{minipage}
    \hfill
    \begin{minipage}[b]{0.49\linewidth}
        \centering
        \includegraphics[width=\linewidth]{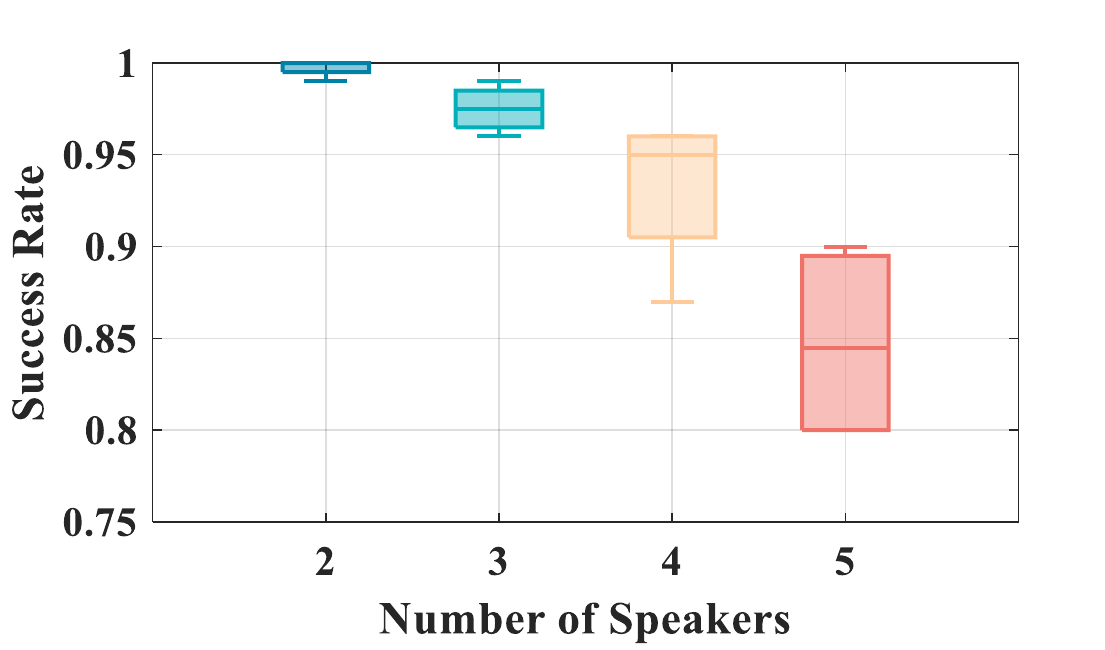}
        \vspace{-0.23in}
        \caption{Speaker \#\ impact.}
        \label{fig:sc_object_layout}
    \end{minipage}
    \vspace{-0.1in}
    \label{fig:layout}
\end{figure}

\noindent \textbf{Impact of Different Object Layout.}
In the real world, objects on a table are often arranged irregularly. To validate our attack's performance under varying object layouts, we consider four additional 3-object layouts, as depicted in Fig. \ref{fig:object_layout_setup}. The three objects are arranged with diverse angular placements, including both acute and obtuse angles.  We describe the speech distinction performance with various layouts in Fig. \ref{fig:sc_object_layout}. It depicts a high average success rate over 0.9 across all 4 layouts in scenarios with $2$ to $4$ speakers. Even in the challenging case of $5$ speakers positioned shoulder-to-shoulder, the median success rate remains above 0.8.

\begin{figure}[H]
    \centering
    \vspace{-0.1in}
        \begin{minipage}[b]{0.58\linewidth}
        \centering
        \includegraphics[width=\linewidth]{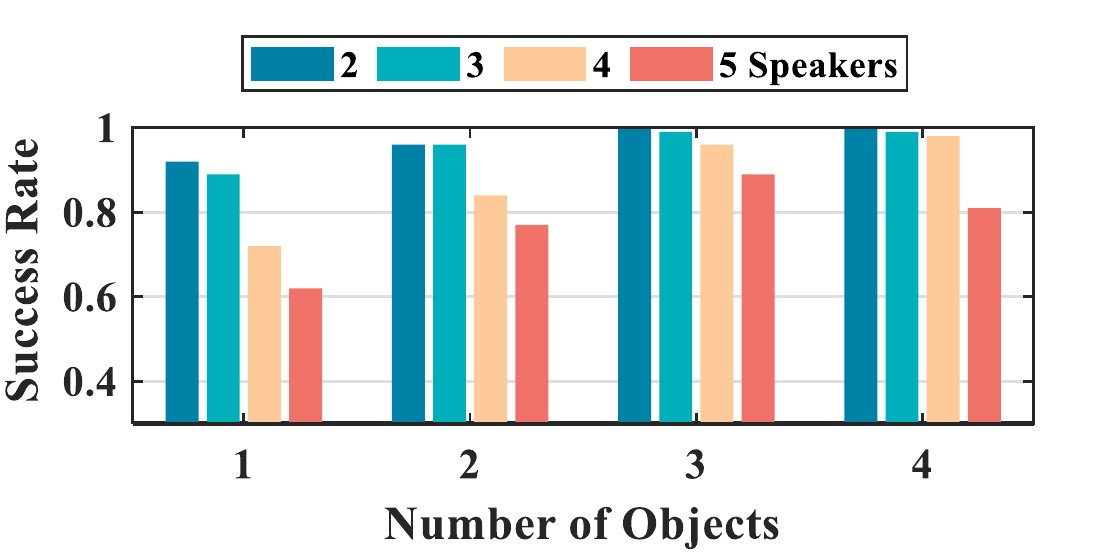}
        \vspace{-0.2in}
        \caption{Object impact. }
        \label{fig:n_objects_acc}
    \end{minipage}
    \hfill
    \begin{minipage}[b]{0.40\linewidth}
        \centering
        \includegraphics[width=\linewidth]{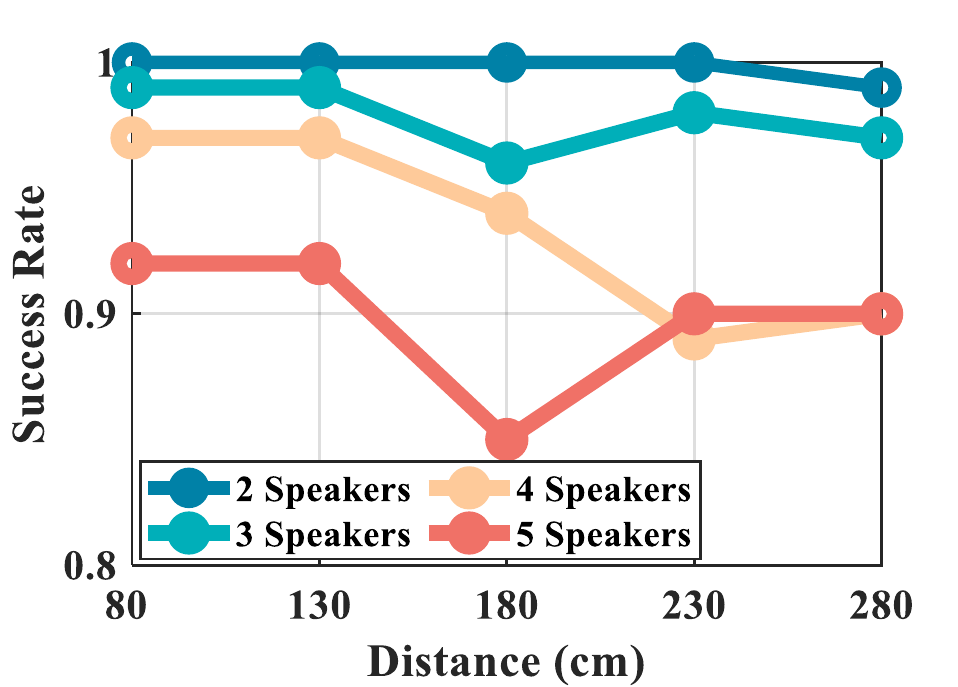}
        \vspace{-0.2in}
        \caption{Distance impact.}
        \label{fig:dis_acc}
    \end{minipage}
    \vspace{-0.3in}
    \label{fig:Res_speaker_layout_short}
\end{figure}

\noindent \textbf{Impact of Number of Objects.}
Fig. \ref{fig:n_objects_acc} shows that distinction performance improves with an increasing number of objects. This trend arises because a greater number of objects provide with more diverse signal sources, thereby enriching the spatial information for eavesdropping. However, the gain plateaus beyond three objects, which indicates that three objects are sufficient to capture the positional information of speakers in our scenario. It implies our attack is practical since it works well when only a few objects are available.
\noindent \textbf{Impact of Sensing Distance.}
In  indoor environments  such as homes and offices, the distance between targets and radar can vary considerably. To assess this effect, we vary the sensing distance between the mmWave radar and the objects from 80 cm to 280 cm. As shown in Fig. \ref{fig:dis_acc}, the success rate exhibits a downward trend as distance increases. Nevertheless, our attack  delivers strong performance: even at the longest distance of 280 cm, it achieves an success rate exceeding 0.9 with both 4 and 5 speakers. This result suggests that our attack remains effective across typical indoor scenarios.

\subsubsection{Performance of Signal Enhancement}  We conduct experiments under different practical setups 

\begin{figure}[H]
    \centering
    \begin{subfigure}[b]{0.23\textwidth}
      \centering
        \includegraphics[width=\linewidth]{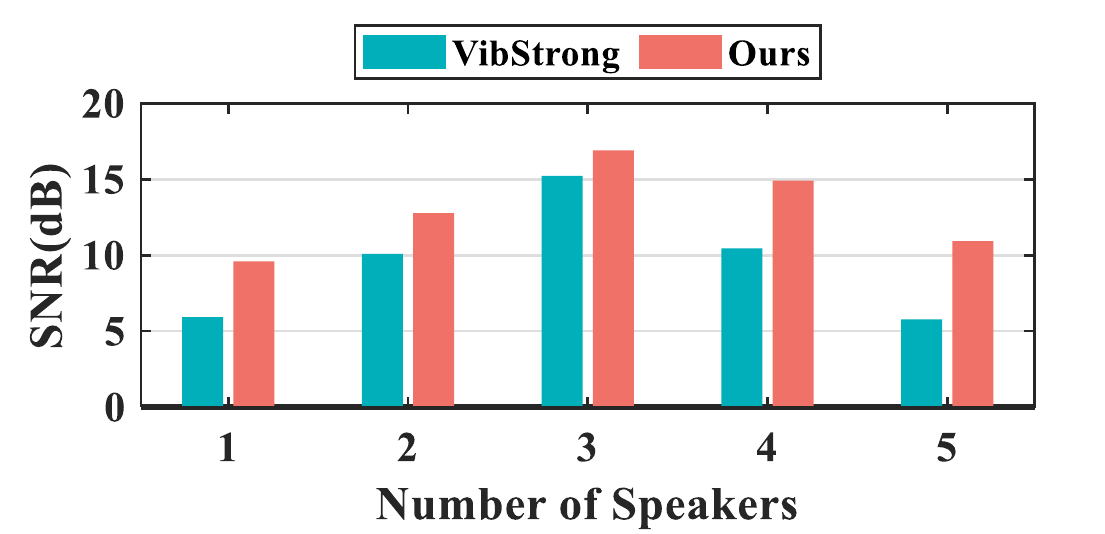}
        \vspace{-0.2in}
        % \caption{SNR}
        \label{fig:material_photo}
    \end{subfigure}
    \begin{subfigure}[b]{0.23\textwidth}
      \centering
        \includegraphics[width=\linewidth]{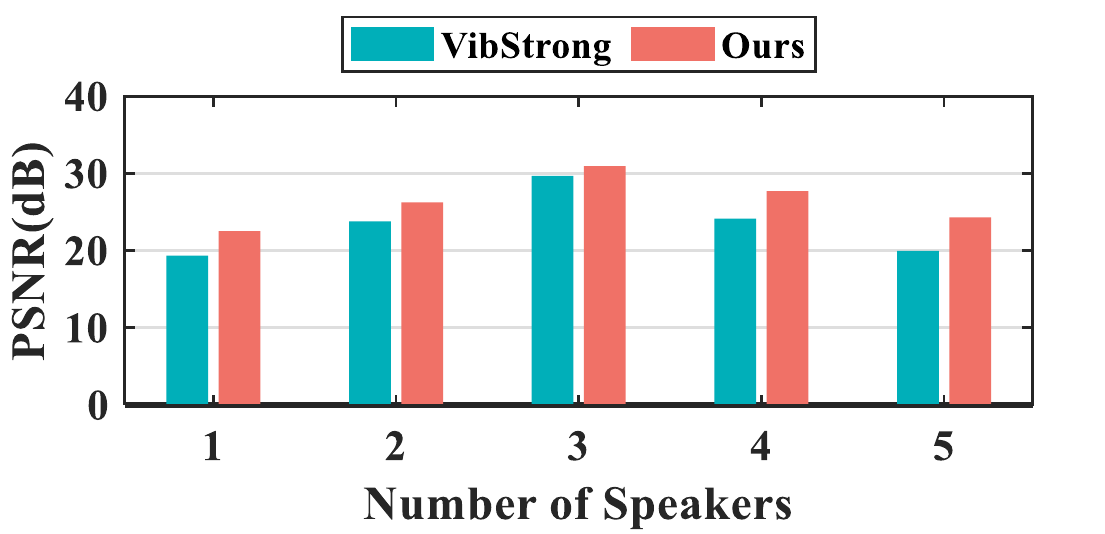}
        \vspace{-0.2in}
        % \caption{PSNR}
        \label{fig:material_result}
    \end{subfigure}
    \caption{Overall speech enhancement performance comparison.}
    \label{fig:se_overall}
    \vspace{-0.1in}
\end{figure}

\noindent\textbf{Overall Performance.}
We evaluate the signal enhancement performance with 5 speakers using the default setup in Fig. \ref{fig:overall_setup}. The results in Fig. \ref{fig:se_overall} indicate that our approach consistently outperforms \textit{VibStrong}, maintaining an average SNR above 10 dB and an average PSNR above 20 dB across scenarios with $2$ to $5$ speakers. We also observe an interesting trend: signal quality initially improves as the number of speakers increases and then declines. This is because, with fewer speakers, they tend to be more spaced out, providing stronger responses from objects. However, as more speakers get involved, their positions shift to less optimal locations (e.g., P2 and P4), resulting in weaker object responses and, consequently, a decline in signal quality. These results demonstrate our attack's capability to effectively enhance signals.
\vspace{-0.1in}

\begin{figure}[H]
\vspace{-0.2in}
    \centering
    \begin{subfigure}[b]{0.48\linewidth}
        \centering
        \includegraphics[width=\linewidth]{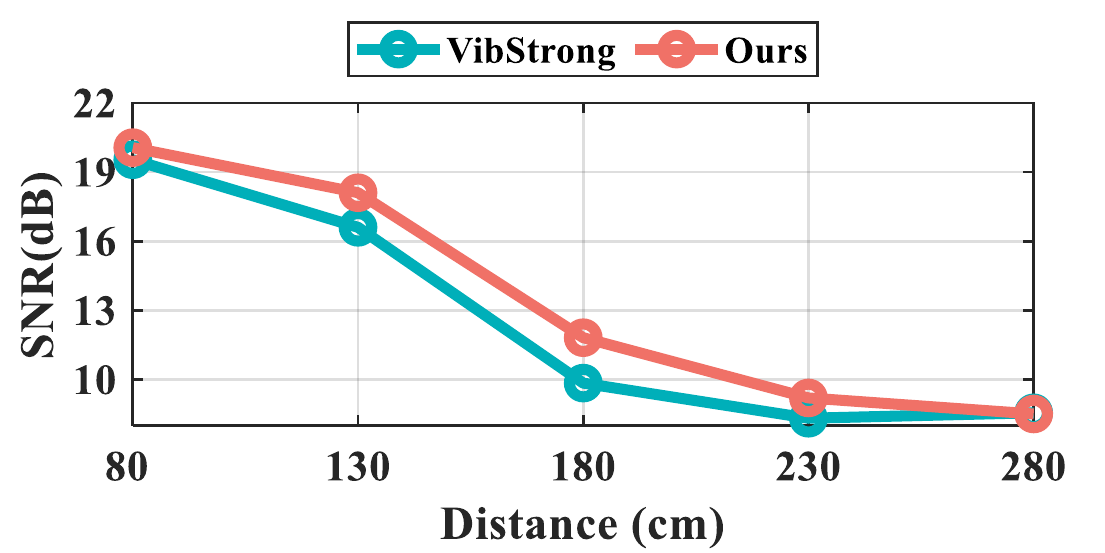}
        \label{fig:snr_dis}
    \end{subfigure}
    \hfill
    \begin{subfigure}[b]{0.48\linewidth}
        \centering
        \includegraphics[width=\linewidth]{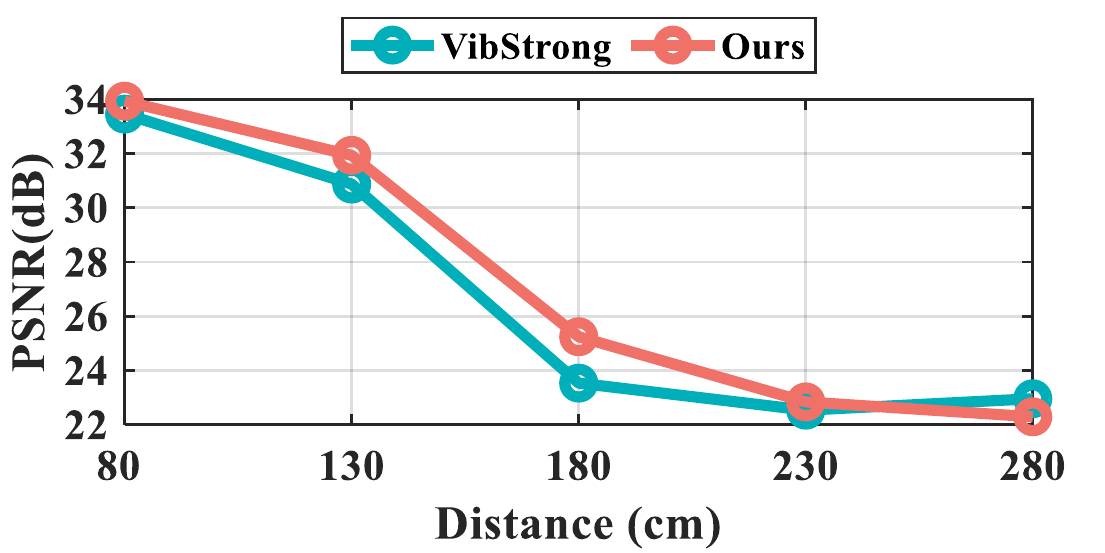}
        \label{fig:psnr_dis}
    \end{subfigure}
    \vspace{-0.2in}
    \caption{Performance under different sensing distances.}
    \vspace{-0.1in}
    \label{fig:se_dis}
\end{figure}
\noindent\textbf{Impact of Sensing Distance.}
We investigate the sensing distance impact by varying the target-to-radar distance from 80 cm to 280 cm. Fig. \ref{fig:se_dis} clearly demonstrates a decrease in performance as the sensing distance increases. This degradation results from the strong attenuation of mmWave reflections at longer distances, which weakens the observable vibrations.
Additionally, the effectiveness of our enhancement scheme diminishes at greater distances. When the distance exceeds 250 cm, the improvement becomes negligible for some objects due to the absence of sufficiently strong vibrations. These results indicates that signal enhancement benefits from the fusion mechanism, which leverages any available strong vibration response. However, when no detectable vibrations are present, the enhancement effect becomes limited.

% , as some objects at further positions are almost unable to capture useful frequency responses, resulting in less effective speech information for signal improvement. \textbf{Therefore, when multiple objects contain detectable vibration responses, our attack demonstrates an improved ability to enhance overall signal quality.}

\subsubsection{Performance of Speech Recovery} 
\begin{figure}[H]
    \vspace{-0.1in}
    \centering
    \includegraphics[width=0.55\linewidth]{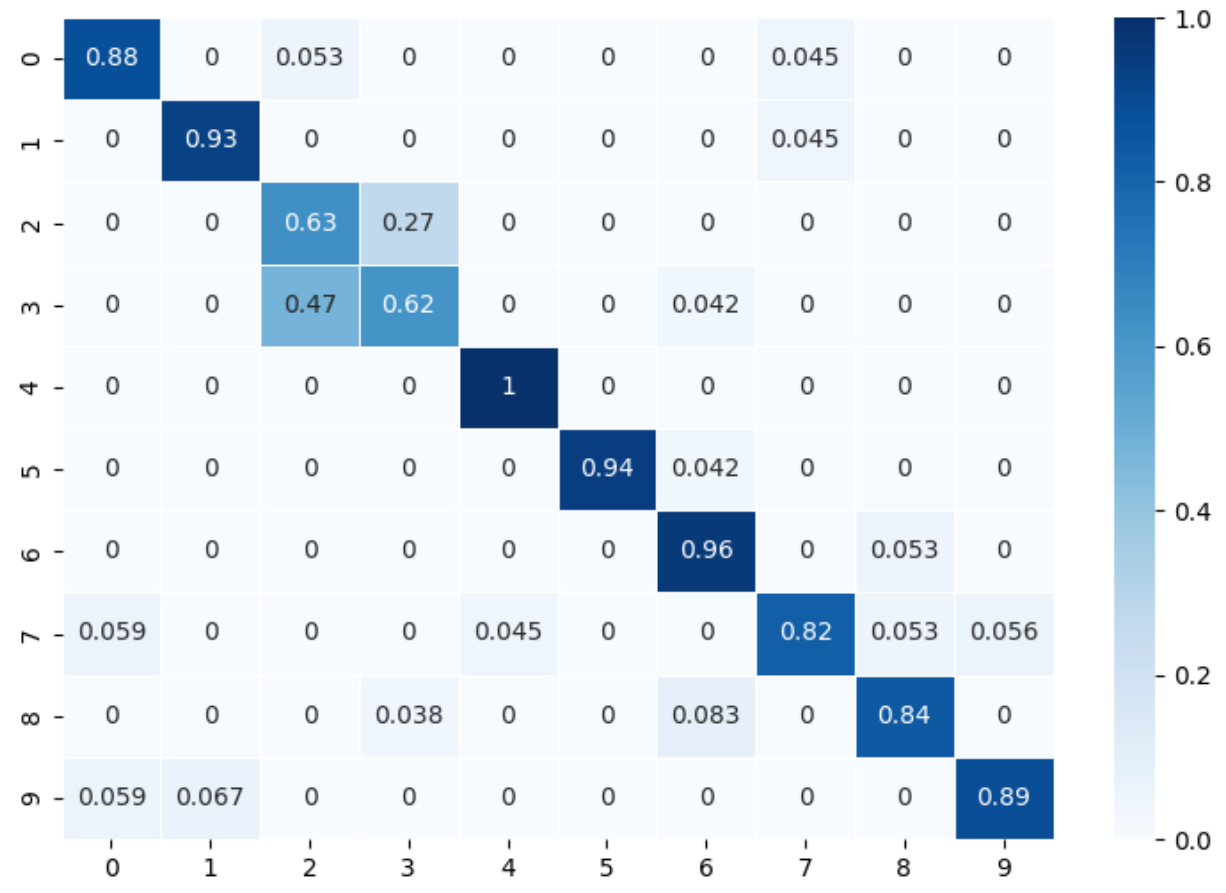}
    \vspace{-0.05in}
    \caption{Speech recovery performance.}
    \label{fig:digit_recognition}
    \vspace{-0.1in}
\end{figure}

To evaluate the effectiveness of our attack in speech recovery, we use digit recognition as a downstream task. We train a four-layer convolutional neural network (CNN) that takes spectrograms generated from our enhanced mmWave signals as input. The signals are collected using the setup in Fig. \ref{fig:p1_layout}. As shown in Fig. \ref{fig:digit_recognition}, the model achieves an overall accuracy of 0.85 across 10 digits. This result confirms that our enhancement method recovers sufficient speech information for accurate recognition.

\subsection{Real-World Case Study}
\begin{figure}[H]
\vspace{-0.1in}
    \centering
    \begin{subfigure}[b]{0.48\linewidth}
        \centering
        \includegraphics[width=0.95\linewidth]{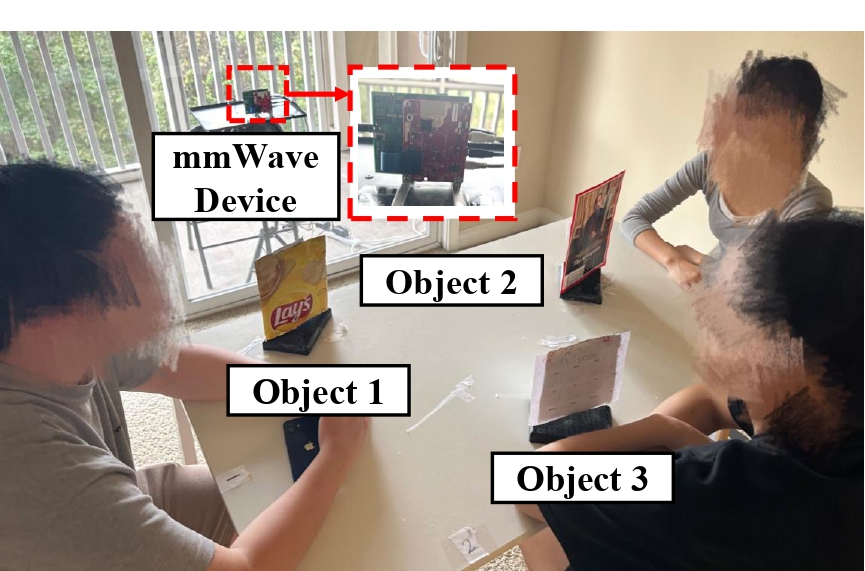}
        \caption{Live human speech scenario.}
        \label{fig:realhuman_setup}
    \end{subfigure}
    \hfill
    \begin{subfigure}[b]{0.48\linewidth}
        \centering
        \includegraphics[width=0.95\linewidth]{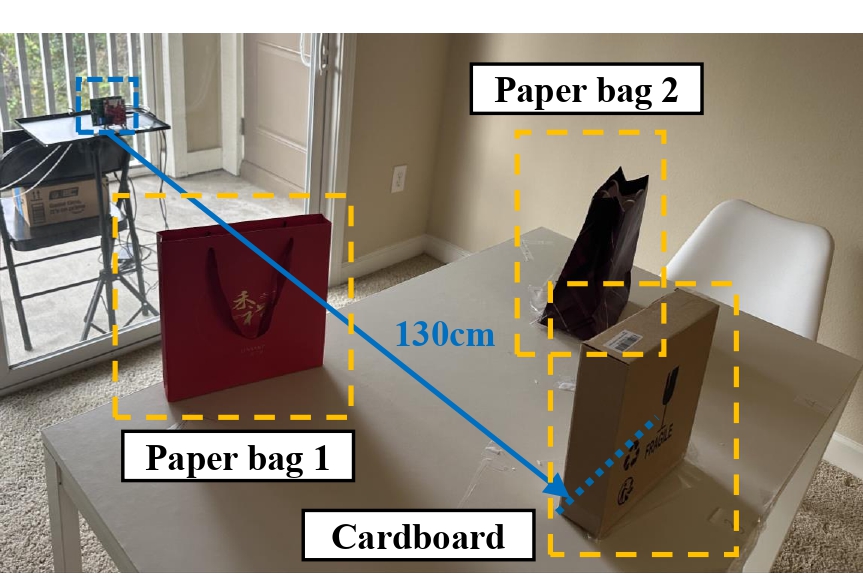}
        \caption{Common objects scenario.}
        \label{fig:sc_material}
    \end{subfigure}
    \vspace{-0.05in}
    \caption{Real-World Case Study Scenarios.}
    \vspace{-0.10in}
    \label{fig:realworld}
\end{figure}

We evaluate our attack under two real-world scenarios: common objects and live human speech, as shown in Fig. \ref{fig:realworld}. In both cases, reflective objects are placed on a table at the same locations used in the default setup.
In the \textit{common object scenario}, we use two paper bags of different sizes and a cardboard box as reflective objects. In the \textit{live human speech scenario}, three volunteers, including one Chinese female P1 and two males (one Chinese P2 and one American P3), act as speakers. They are seated naturally along three edges of a table, while three tinfoil sheets serve as reflective objects.
\begin{table}[H]
\vspace{-0.1in}
    \centering
    \begin{tabular}{l|cccc}
        \specialrule{1.5pt}{1pt}{1pt}
        \textbf{\# of Speakers} & 2 & 3 & 4 & 5 \\
        \specialrule{1pt}{1pt}{1pt}
        \textbf{Success Rate} & 1.0 & 1.0 & 0.99 & 0.95 \\
        \specialrule{1.5pt}{1pt}{1pt}
    \end{tabular}
    \vspace{-0.05in}
    \caption{Speaker distinction with common object scenario.}
    \vspace{-0.10in}
    \label{tab:common_objects_acc}
\end{table}
\subsubsection{Performance of Speaker Distinction}
Table \ref{tab:common_objects_acc} reports the speaker distinction performance under the \textit{common objects scenario} with 2 to 5 speakers. The results show strong and consistent performance: the success rate remains above 0.99 for 2–4 speakers, and only slightly decreases to 0.95 in the more challenging 5-speaker case.  This indicates that, even when speech-induced vibrations are reflected by irregularly shaped and lightweight objects, our attack can reliably extract frequency responses for accurate speaker distinction. This highlights the practical applicability of our approach in realistic environments with non-ideal and unstructured reflectors.

In the \textit{live human speech scenario}, three volunteers are asked to  speak digits from 0 to 9, each repeating them 20 times. We collect the mmWave data and evaluate speaker distinction performance.
As shown in Table \ref{tab:human_acc}, the success rate consistently reaches 0.99 across all speaker arrangements. It demonstrates that our attack remains effective despite variations in speaker identity, such as gender and nationality. These findings further validate the reliability of our method in distinguishing individuals during real-world live speech. 
\begin{table}[H]
\vspace{-0.05in}
    \centering
    \begin{tabular}{l|cccc}
        \specialrule{1.5pt}{1pt}{1pt}
        \textbf{Speaker Layout} & P1P2 & P2P3 & P1P3 & P1 P2 P3 \\
        \specialrule{1pt}{1pt}{1pt}
        \textbf{Success Rate} & 0.99 & 0.99 & 0.99 & 0.99 \\
        \specialrule{1.5pt}{1pt}{1pt}
    \end{tabular}
    % \vspace{-0.05in}
    \caption{Speaker distinction under live human speech.}
    \vspace{-0.1in}
    \label{tab:human_acc}
\end{table}

\subsubsection{Performance of Signal Enhancement}
\begin{figure}[H]
\vspace{-0.1in}
    \centering
    \begin{subfigure}[b]{0.48\linewidth}
        \centering
        \includegraphics[width=\linewidth]{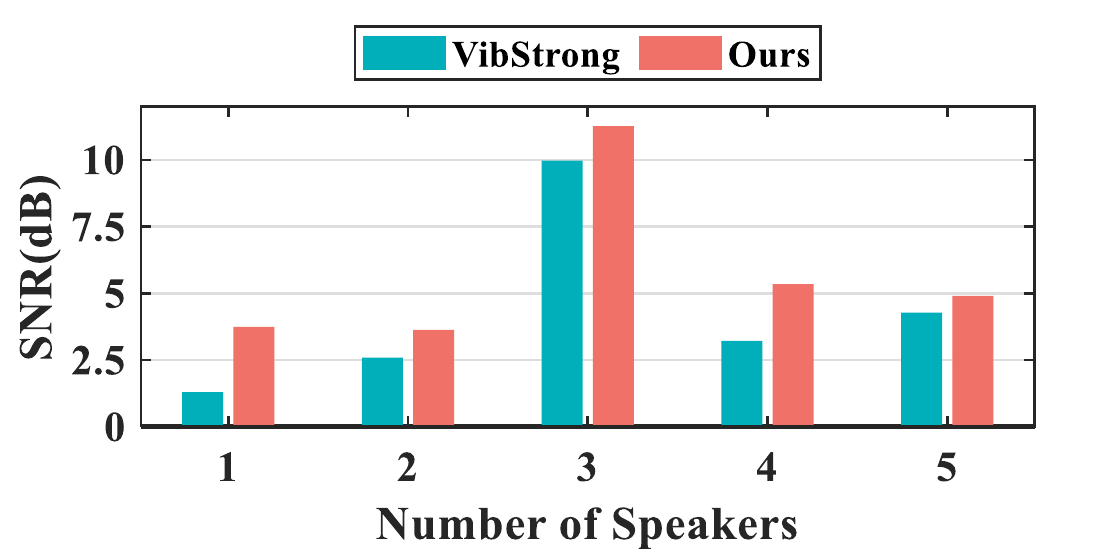}
        % \caption{}
        \label{fig:snr_material}
    \end{subfigure}
    \hfill
    \begin{subfigure}[b]{0.48\linewidth}
        \centering
        \includegraphics[width=\linewidth]{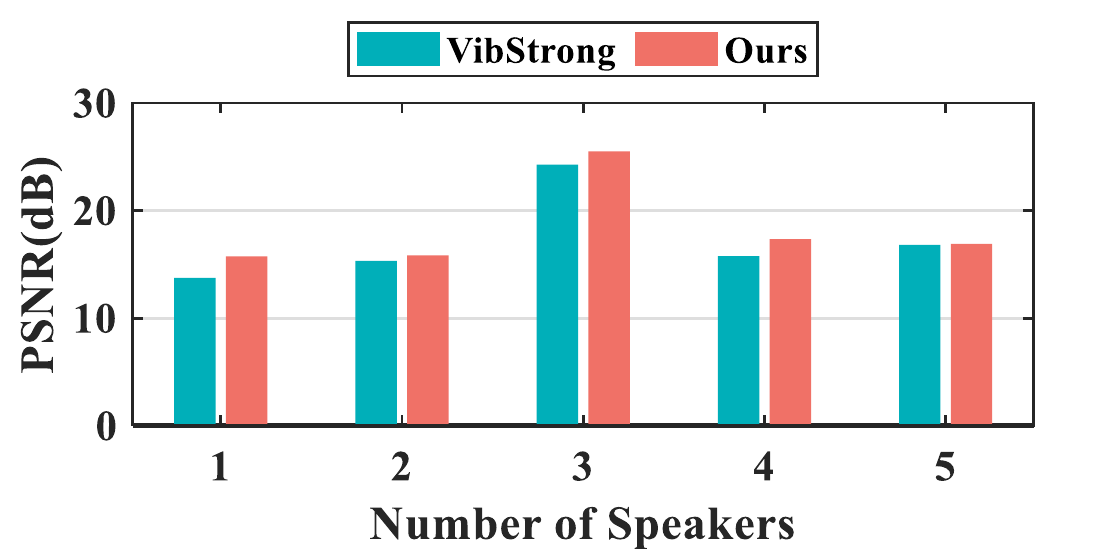}
        % \caption{}
        \label{fig:psnr_material}
    \end{subfigure}
    \vspace{-0.2in}
    \caption{Signal enhancement under common object scenario.}
    \vspace{-0.1in}
    \label{fig:se_common}
\end{figure}

As shown in Fig. \ref{fig:se_common}, the enhancement performance under the \textit{common object scenario} improves significantly as the number of speakers increases to three. This gain is primarily attributed to the third speaker, who is positioned directly in front of the cardboard, which provides a stronger frequency response than the paper bags. These results suggest that our attack can achieve reliable enhancement performance even with common household objects, and performs better when reflectors exhibit more pronounced vibration responses.

\begin{figure}[H]
\vspace{-0.1in}
    \centering
    \begin{subfigure}[b]{0.48\linewidth}
        \centering
        \includegraphics[width=\linewidth]{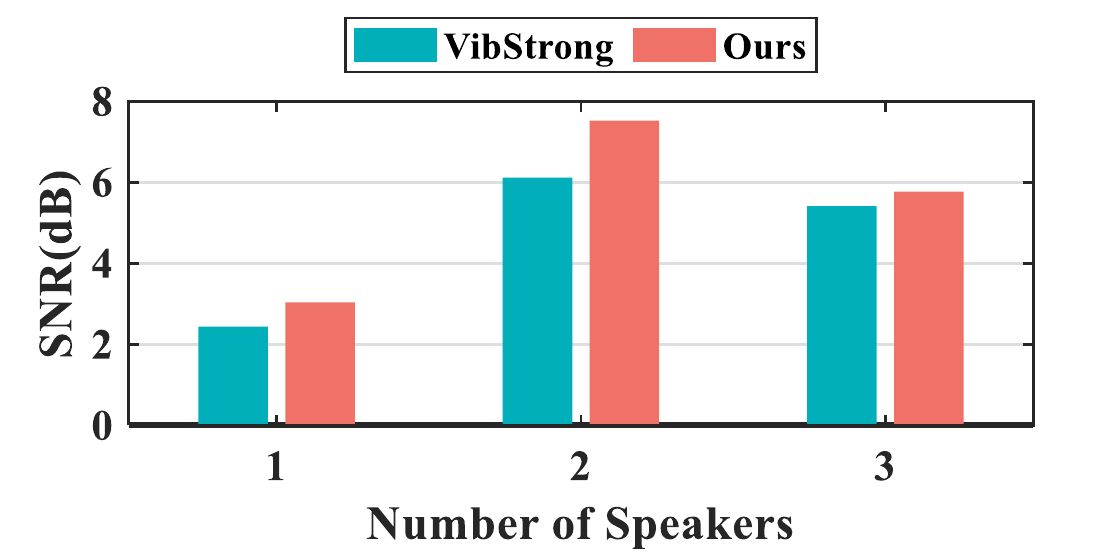}
        \label{fig:snr_realhuman}
    \end{subfigure}
    \hfill
    \begin{subfigure}[b]{0.48\linewidth}
        \centering
        \includegraphics[width=\linewidth]{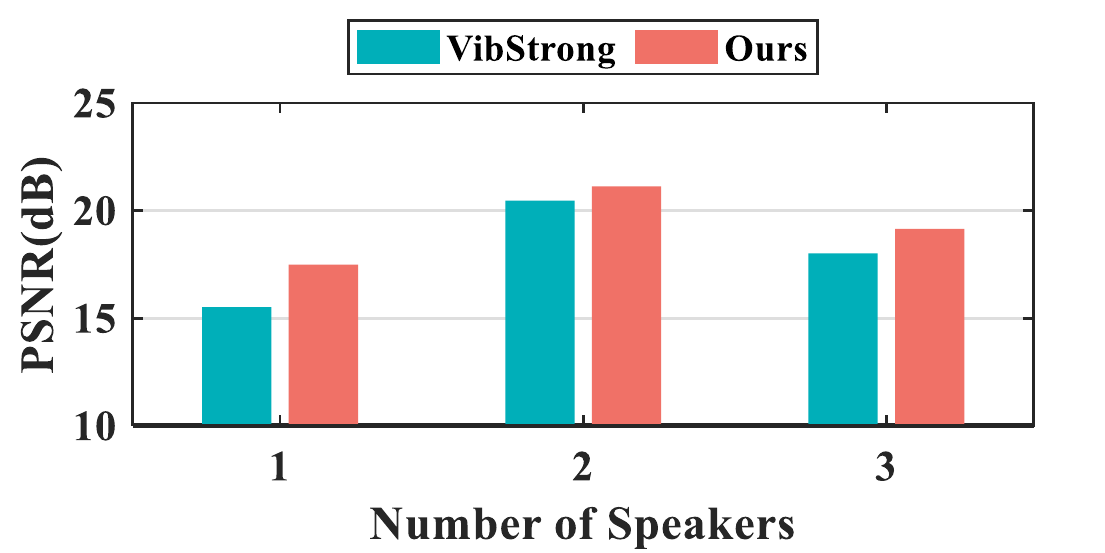}
        \label{fig:psnr_realhuman}
    \end{subfigure}
    \vspace{-0.2in}
    \caption{Signal enhancement under live human speech.}
    \vspace{-0.1in}
    \label{fig:se_realhuman}
\end{figure}

A similar trend is observed under the \textit{live human speech scenario}. As shown in Fig. \ref{fig:se_realhuman}, when the number of participants increases to 2 or 3, the enhanced signal quality substantially improves. This is because the additional participants, P2 and P3, are both males, with lower vocal frequencies that produce stronger low-frequency energy, leading to more pronounced responses on the objects. Nevertheless, our attack consistently enhances speech signals across varying speaking styles and speaker genders.

\section{Related Work}
\textbf{Motion sensor-based sound eavesdropping.} Motion sensors can  detect motion events caused by speech, showing the feasibility of acoustic eavesdropping \cite{cao2023can, zhang2023spy, hu2022accear}. The motion caused by the speech-induced vibrations leads to changes in capacitance, resistance, or piezoelectric elements within motion sensors. Those changes are then converted into electrical signals, which are leveraged to launch attacks. For example, the attacker uses the built-in accelerometer in a smartphone to eavesdrop on the speaker \cite{ba2020learning}. 
% For example, The relative motion resulted from speech-induced minute vibrations leads to changes in the capacitance, resistance, or piezoelectric elements within accelerometers of the smartphone. 
% \xz{Attacker can eavesdrop with the speaker of the same smartphone \cite{ba2020learning}}. 
Besides, via reading data from built-in gyroscope in a smartphone, the attacker can recover the speech from the speaker \cite{michalevsky2014gyrophone}. However, these approaches require access to the smartphone to obtain sensor data and have low sampling rates as well. In contrast, our attack employs a mmWave radar for sound eavesdropping, which can be deployed in inconspicuous positions. Its high vibration sampling rate, approximately 10 kHz, allows for more accurate capture of minute vibrations caused by speech, improving the feasibility and effectiveness of sound eavesdropping.

\textbf{Vision-based sound eavesdropping.} High resolution vision sensors with high sampling rates can be deployed to perform sound eavesdropping via detecting minute vibrations. For instance, the high-speed camera is utilized to capture videos of objects and extract minute vibrations on the object surface induced by speech \cite{davis2014visual}. For the indoor scenario investigated in \cite{sami2020spying}, the vacuum cleaning robots are equipped with LiDAR sensors to detect vibrations induced on nearby objects to recover speech. For long-range attacks, Ben et al. in \cite{nassi2020lamphone} analyze a hanging light bulb’s vibration response captured by remote electro-optical sensors. Similarly, EchoLight uses a photodetector behind a telescope to detect light changes caused by speech-induced vibrations on everyday objects \cite{zhang2024echolight}.
However, vision-based approaches perform poorly in low-light conditions and are easily obstructed. In comparison, the mmWave signals in our attack can penetrate opaque obstacles, thus avoiding these limitations. 

\textbf{RF-based sound eavesdropping.} Radio frequency signals, such as WiFi \cite{wei2015acoustic, hu2023password}, RFID \cite{wang2021thru, chen2024rfspy}, and mmWave \cite{wang2022wavesdropper, hu2022milliear, shi2023privacy, wang2024vibspeech, xu2024mmear}, are popular in sound eavesdropping. Due to the penetrating capabilities of RF signals, they are deployed for undetectable attacks outside the room. For example, attaching battery-less
commercial off-the-shelf RFID tags on everyday objects can achieve through-the-wall sound eavesdropping \cite{wang2021thru}. mmWave-based sound eavesdropping approaches have gained significant attention recently because of its higher movement resolution. It allows speech recovery through the audio source itself (human throat \cite{wang2022wavesdropper} or loudspeaker surface \cite{hu2022milliear, feng2023mmeavesdropper})  or nearby everyday objects \cite{hu2023mmecho, shi2023privacy}. The latter approach detects surface vibrations induced by sound waves on nearby objects, enabling speech recovery even when the original audio source is not directly accessible. However, current approaches mainly focus on using only the nearest object signal for sound eavesdropping. They neglect the fact that other object signals also contain valuable speech information. In this work, we leverage spatial diversity from multiple objects for speech distinction and signal enhancement.

\section{Conclusion}
This paper presents a novel attack that enables remote eavesdropping on in-person conversations via mmWave sensing. 
We develop a noise-robust signal processing pipeline to extract speaker-specific vibration patterns for speaker distinction.
 On the other hand, we design a deep learning framework that enhances signal quality by fusing multi-object vibration data. Our extensive experiments demonstrate that the proposed attack achieves a high speaker distinction success rate and remains robust across diverse and practical setups. These findings highlight emerging privacy risks associated with passive sensing in shared environments.

% In this paper, we present a novel attack for in-person conversation eavesdropping via mmWave sensing. Our attack exploits the spatial diversity brought by multiple objects. We distinguish speakers by detecting frequency response differences of speech-induced vibrations on objects. By combining signals from multiple objects, we further enhance the recovered signal quality. Our extensive experimental results validate the effectiveness and robustness of our attack in real-world scenarios.

\bibliographystyle{IEEEtran}
\bibliography{main}

@inproceedings{ba2020learning,
  title={Learning-based Practical Smartphone Eavesdropping with Built-in Accelerometer.},
  author={Ba, Zhongjie and Zheng, Tianhang and Zhang, Xinyu and Qin, Zhan and Li, Baochun and Liu, Xue and Ren, Kui},
  booktitle={NDSS},
  volume={2020},
  pages={1--18},
  year={2020}
}

@inproceedings{michalevsky2014gyrophone,
  title={Gyrophone: Recognizing speech from gyroscope signals},
  author={Michalevsky, Yan and Boneh, Dan and Nakibly, Gabi},
  booktitle={23rd USENIX Security Symposium (USENIX Security 14)},
  pages={1053--1067},
  year={2014}
}

@inproceedings{cao2023can,
  title={I Can Hear You Without a Microphone: Live Speech Eavesdropping From Earphone Motion Sensors},
  author={Cao, Yetong and Li, Fan and Chen, Huijie and Liu, Xiaochen and Duan, Chunhui and Wang, Yu},
  booktitle={IEEE INFOCOM 2023-IEEE Conference on Computer Communications},
  pages={1--10},
  year={2023},
  organization={IEEE}
}

@article{davis2014visual,
  title={The visual microphone: Passive recovery of sound from video},
  author={Davis, Abe and Rubinstein, Michael and Wadhwa, Neal and Mysore, Gautham J and Durand, Fredo and Freeman, William T},
  year={2014},
  publisher={Association for Computing Machinery (ACM)}
}

@inproceedings{wei2015acoustic,
  title={Acoustic eavesdropping through wireless vibrometry},
  author={Wei, Teng and Wang, Shu and Zhou, Anfu and Zhang, Xinyu},
  booktitle={Proceedings of the 21st Annual International Conference on Mobile Computing and Networking},
  pages={130--141},
  year={2015}
}

@article{wang2021thru,
  title={Thru-the-wall eavesdropping on loudspeakers via RFID by capturing sub-mm level vibration},
  author={Wang, Chuyu and Xie, Lei and Lin, Yuancan and Wang, Wei and Chen, Yingying and Bu, Yanling and Zhang, Kai and Lu, Sanglu},
  journal={Proceedings of the ACM on Interactive, Mobile, Wearable and Ubiquitous Technologies},
  volume={5},
  number={4},
  pages={1--25},
  year={2021},
  publisher={ACM New York, NY, USA}
}

@inproceedings{wang2022mmeve,
  title={mmEve: eavesdropping on smartphone's earpiece via COTS mmWave device},
  author={Wang, Chao and Lin, Feng and Liu, Tiantian and Zheng, Kaidi and Wang, Zhibo and Li, Zhengxiong and Huang, Ming-Chun and Xu, Wenyao and Ren, Kui},
  booktitle={Proceedings of the 28th Annual International Conference on Mobile Computing And Networking},
  pages={338--351},
  year={2022}
}

@inproceedings{jiang2020mmvib,
  title={mmVib: micrometer-level vibration measurement with mmwave radar},
  author={Jiang, Chengkun and Guo, Junchen and He, Yuan and Jin, Meng and Li, Shuai and Liu, Yunhao},
  booktitle={Proceedings of the 26th Annual International Conference on Mobile Computing and Networking},
  pages={1--13},
  year={2020}
}

@article{anguera2012speaker,
  title={Speaker diarization: A review of recent research},
  author={Anguera, Xavier and Bozonnet, Simon and Evans, Nicholas and Fredouille, Corinne and Friedland, Gerald and Vinyals, Oriol},
  journal={IEEE Transactions on audio, speech, and language processing},
  volume={20},
  number={2},
  pages={356--370},
  year={2012},
  publisher={IEEE}
}

@inproceedings{wang2018speaker,
  title={Speaker diarization with LSTM},
  author={Wang, Quan and Downey, Carlton and Wan, Li and Mansfield, Philip Andrew and Moreno, Ignacio Lopz},
  booktitle={2018 IEEE International conference on acoustics, speech and signal processing (ICASSP)},
  pages={5239--5243},
  year={2018},
  organization={IEEE}
}

@inproceedings{hu2023mmecho,
  title={mmecho: A mmwave-based acoustic eavesdropping method},
  author={Hu, Pengfei and Li, Wenhao and Spolaor, Riccardo and Cheng, Xiuzhen},
  booktitle={Proceedings of the ACM Turing Award Celebration Conference-China 2023},
  pages={138--140},
  year={2023}
}

@inproceedings{feng2023mmeavesdropper,
  title={mmeavesdropper: Signal augmentation-based directional eavesdropping with mmwave radar},
  author={Feng, Yiwen and Zhang, Kai and Wang, Chuyu and Xie, Lei and Ning, Jingyi and Chen, Shijia},
  booktitle={IEEE INFOCOM 2023-IEEE Conference on Computer Communications},
  pages={1--10},
  year={2023},
  organization={IEEE}
}

@book{jaramillo2014architectural,
  title={Architectural acoustics},
  author={Jaramillo, Ana and Steel, Chris},
  year={2014},
  publisher={Routledge}
}

@inbook{inbook,
author = {Arellano-Castro, Rocio and Marin, María and Cros, Anne},
year = {2011},
month = {09},
pages = {},
title = {Forced Oscillations of a Membrane},
isbn = {3642179584, 9783642179587},
doi = {10.1007/978-3-642-17958-7_35}
}

@article{zhang2022ambiear,
  title={Ambiear: mmwave based voice recognition in nlos scenarios},
  author={Zhang, Jia and Zhou, Yinian and Xi, Rui and Li, Shuai and Guo, Junchen and He, Yuan},
  journal={Proceedings of the ACM on Interactive, Mobile, Wearable and Ubiquitous Technologies},
  volume={6},
  number={3},
  pages={1--25},
  year={2022},
  publisher={ACM New York, NY, USA}
}

@inproceedings{jiang2021sense,
  title={Sense me on the ride: Accurate mobile sensing over a LoRa backscatter channel},
  author={Jiang, Haotian and Zhang, Jiacheng and Guo, Xiuzhen and He, Yuan},
  booktitle={Proceedings of the 19th ACM Conference on Embedded Networked Sensor Systems},
  pages={125--137},
  year={2021}
}

@article{westfall2014kurtosis,
  title={Kurtosis as peakedness, 1905--2014. RIP},
  author={Westfall, Peter H},
  journal={The American Statistician},
  volume={68},
  number={3},
  pages={191--195},
  year={2014},
  publisher={Taylor \& Francis}
}

@inproceedings{shi2023privacy,
  title={Privacy Leakage via Speech-induced Vibrations on Room Objects through Remote Sensing based on Phased-MIMO},
  author={Shi, Cong and Zhang, Tianfang and Xu, Zhaoyi and Li, Shuping and Gao, Donglin and Li, Changming and Petropulu, Athina and Wu, Chung-Tse Michael and Chen, Yingying},
  booktitle={Proceedings of the 2023 ACM SIGSAC Conference on Computer and Communications Security},
  pages={75--89},
  year={2023}
}

@article{chang2007constrained,
  title={Constrained least-squares optimization for robust estimation of center of rotation},
  author={Chang, Lillian Y and Pollard, Nancy S},
  journal={Journal of biomechanics},
  volume={40},
  number={6},
  pages={1392--1400},
  year={2007},
  publisher={Elsevier}
}

@article{reynolds2009gaussian,
  title={Gaussian mixture models.},
  author={Reynolds, Douglas A and others},
  journal={Encyclopedia of biometrics},
  volume={741},
  number={659-663},
  year={2009},
  publisher={Berlin, Springer}
}

@inproceedings{erdogan2016improved,
  title={Improved mvdr beamforming using single-channel mask prediction networks.},
  author={Erdogan, Hakan and Hershey, John R and Watanabe, Shinji and Mandel, Michael I and Le Roux, Jonathan},
  booktitle={Interspeech},
  pages={1981--1985},
  year={2016}
}

@inproceedings{sami2020spying, 

  title={Spying with your robot vacuum cleaner: eavesdropping via lidar sensors}, 

  author={Sami, Sriram and Dai, Yimin and Tan, Sean Rui Xiang and Roy, Nirupam and Han, Jun}, 

  booktitle={Proceedings of the 18th Conference on Embedded Networked Sensor Systems}, 

  pages={354--367}, 

  year={2020} 

}

@article{nassi2020lamphone, 

  title={Lamphone: Real-time passive sound recovery from light bulb vibrations}, 

  author={Nassi, Ben and Pirutin, Yaron and Shamir, Adi and Elovici, Yuval and Zadov, Boris}, 

  journal={Cryptology ePrint Archive}, 

  year={2020} 

}

@article{wang2022wavesdropper,
  title={Wavesdropper: Through-wall word detection of human speech via commercial mmWave devices},
  author={Wang, Chao and Lin, Feng and Ba, Zhongjie and Zhang, Fan and Xu, Wenyao and Ren, Kui},
  journal={Proceedings of the ACM on Interactive, Mobile, Wearable and Ubiquitous Technologies},
  volume={6},
  number={2},
  pages={1--26},
  year={2022},
  publisher={ACM New York, NY, USA}
}

@inproceedings{hu2022milliear,
  title={Milliear: Millimeter-wave acoustic eavesdropping with unconstrained vocabulary},
  author={Hu, Pengfei and Ma, Yifan and Santhalingam, Panneer Selvam and Pathak, Parth H and Cheng, Xiuzhen},
  booktitle={IEEE INFOCOM 2022-IEEE Conference on Computer Communications},
  pages={11--20},
  year={2022},
  organization={IEEE}
}

@article{audiomnist2023,
    title = {AudioMNIST: Exploring Explainable Artificial Intelligence for audio analysis on a simple benchmark},
    journal = {Journal of the Franklin Institute},
    year = {2023},
    issn = {0016-0032},
    doi = {https://doi.org/10.1016/j.jfranklin.2023.11.038},
    url = {https://www.sciencedirect.com/science/article/pii/S0016003223007536},
    author = {Sören Becker and Johanna Vielhaben and Marcel Ackermann and Klaus-Robert Müller and Sebastian Lapuschkin and Wojciech Samek},
}

@inproceedings{panayotov2015librispeech,
  title={Librispeech: an asr corpus based on public domain audio books},
  author={Panayotov, Vassil and Chen, Guoguo and Povey, Daniel and Khudanpur, Sanjeev},
  booktitle={2015 IEEE international conference on acoustics, speech and signal processing (ICASSP)},
  pages={5206--5210},
  year={2015},
  organization={IEEE}
}

@article{gavin2019levenberg,
  title={The Levenberg-Marquardt algorithm for nonlinear least squares curve-fitting problems},
  author={Gavin, Henri P},
  journal={Department of Civil and Environmental Engineering Duke University August},
  volume={3},
  year={2019}
}

@book{kinsler2000fundamentals,
  title={Fundamentals of acoustics},
  author={Kinsler, Lawrence E and Frey, Austin R and Coppens, Alan B and Sanders, James V},
  year={2000},
  publisher={John wiley \& sons}
}

@article{zhang2023spy,
  title={I spy you: Eavesdropping continuous speech on smartphones via motion sensors},
  author={Zhang, Shijia and Liu, Yilin and Gowda, Mahanth},
  journal={Proceedings of the ACM on Interactive, Mobile, Wearable and Ubiquitous Technologies},
  volume={6},
  number={4},
  pages={1--31},
  year={2023},
  publisher={ACM New York, NY, USA}
}

@article{iovescu2017fundamentals,
  title={The fundamentals of millimeter wave sensors},
  author={Iovescu, Cesar and Rao, Sandeep},
  journal={Texas Instruments},
  pages={1--8},
  year={2017}
}

@misc{a2024_logitech,
  title = {Logitech},
  url = {https://www.ultimateears.com/en-us/shop/p/wonderboom-3-bluetooth.984-001807},
  urldate = {2024-11-14},
  year = {2024},
  organization = {Logitech}
}

@inproceedings{hu2022accear,
  title={Accear: Accelerometer acoustic eavesdropping with unconstrained vocabulary},
  author={Hu, Pengfei and Zhuang, Hui and Santhalingam, Panneer Selvam and Spolaor, Riccardo and Pathak, Parth and Zhang, Guoming and Cheng, Xiuzhen},
  booktitle={2022 IEEE Symposium on Security and Privacy (SP)},
  pages={1757--1773},
  year={2022},
  organization={IEEE}
}

@inproceedings{wang2024vibspeech,
  title={$\{$VibSpeech$\}$: Exploring Practical Wideband Eavesdropping via Bandlimited Signal of Vibration-based Side Channel},
  author={Wang, Chao and Lin, Feng and Yan, Hao and Wu, Tong and Xu, Wenyao and Ren, Kui},
  booktitle={33rd USENIX Security Symposium (USENIX Security 24)},
  pages={3997--4014},
  year={2024}
}

@inproceedings{hu2023password,
  title={Password-Stealing without Hacking: Wi-Fi Enabled Practical Keystroke Eavesdropping},
  author={Hu, Jingyang and Wang, Hongbo and Zheng, Tianyue and Hu, Jingzhi and Chen, Zhe and Jiang, Hongbo and Luo, Jun},
  booktitle={Proceedings of the 2023 ACM SIGSAC Conference on Computer and Communications Security},
  pages={239--252},
  year={2023}
}

@article{siddiq2018phase,
  title={Phase noise in FMCW radar systems},
  author={Siddiq, Kashif and Hobden, Mervyn K and Pennock, Steve R and Watson, Robert J},
  journal={IEEE Transactions on Aerospace and Electronic Systems},
  volume={55},
  number={1},
  pages={70--81},
  year={2018},
  publisher={IEEE}
}

@article{boll1979suppression,
  title={Suppression of acoustic noise in speech using spectral subtraction},
  author={Boll, Steven},
  journal={IEEE Transactions on acoustics, speech, and signal processing},
  volume={27},
  number={2},
  pages={113--120},
  year={1979},
  publisher={IEEE}
}

@inproceedings{zhang2024echolight,
  title={EchoLight: Sound eavesdropping based on ambient light reflection},
  author={Zhang, Guoming and Xiang, Zhijie and Fu, Heqiang and Yang, Yanni and Hu, Pengfei},
  booktitle={IEEE INFOCOM 2024-IEEE Conference on Computer Communications},
  pages={341--350},
  year={2024},
  organization={IEEE}
}

@inproceedings{chen2024rfspy,
  title={RFSpy: Eavesdropping on online conversations with out-of-vocabulary words by sensing metal coil vibration of headsets leveraging RFID},
  author={Chen, Yunzhong and Yu, Jiadi and Chen, Yingying and Kong, Linghe and Zhu, Yanmin and Chen, Yi-Chao},
  booktitle={Proceedings of the 22nd Annual International Conference on Mobile Systems, Applications and Services},
  pages={169--182},
  year={2024}
}

@inproceedings{xu2024mmear,
  title={mmEar: Push the limit of COTS mmWave eavesdropping on headphones},
  author={Xu, Xiangyu and Chen, Yu and Ling, Zhen and Lu, Li and Luo, Junzhou and Fu, Xinwen},
  booktitle={IEEE INFOCOM 2024-IEEE Conference on Computer Communications},
  pages={351--360},
  year={2024},
  organization={IEEE}
}

\end{document}